\documentclass{article}
\usepackage[T1]{fontenc}
\usepackage{csquotes}
\usepackage{amsmath,amssymb,amsthm}
\usepackage{mathtools}
\usepackage[table]{xcolor}
\usepackage{array,booktabs}
\usepackage{tikz}
\usepackage{braket}
\usepackage{microtype}
\usepackage{xurl}
\usepackage{hyperref}
\usepackage[letterpaper]{geometry}

\theoremstyle{definition}
\newtheorem{definition}{Definition}

\theoremstyle{plain}
\newtheorem{lemma}{Lemma}
\newtheorem{theorem}{Theorem}
\newtheorem{proposition}{Proposition}
\newtheorem{corollary}{Corollary}

\theoremstyle{remark}
\newtheorem{remark}{Remark}

\usepackage{biblatex}
\newcommand{\bigO}[1]{\mathcal{O}\!\left(#1\right)}
\newcommand{\C}{\mathbb{C}}

\newcommand{\Z}{\mathbb{Z}}

\newcommand{\lcm}{\operatorname{lcm}}

\newcommand{\classP}{\mathrm{P}}
\newcommand{\classNP}{\mathrm{NP}}
\newcommand{\void}[1]{}

\usetikzlibrary{shapes,arrows.meta,positioning,calc,intersections,shapes.geometric}

\definecolor{BEgen}{RGB}{214,244,214}   
\definecolor{BEmat}{RGB}{255,249,196}   
\definecolor{BEpad}{RGB}{255,230,179}   
\definecolor{BEunit}{RGB}{41,121,255}   
\definecolor{BEcoreA}{RGB}{255,153,51}  

\tikzset{
  generaltensor/.style={
    draw=black!50,
    line width=0.6pt,
    circle,
    minimum width=10mm,
    fill=BEgen!85,
  },
  directedtensor/.style={
    draw=black!50,
    rounded rectangle,
    minimum width=20mm,
    minimum height=10mm,
    fill=BEgen!50!BEmat, 
    inner sep=1pt
  },
  matrix/.style={
    draw=black!55,
    line width=0.6pt,
    rectangle,
    minimum width=20mm,
    minimum height=10mm,
    fill=BEmat,
    rounded corners=1pt
  },
  square_matrix/.style={draw=black!60, fill=BEpad},
  unitary/.style={draw=BEunit!60!black, fill=BEunit!12, regular polygon,regular polygon sides=3,minimum size=12mm,inner sep=1pt},
  isometry/.style={draw=BEunit!60!black, fill=BEunit!10, isosceles triangle, isosceles triangle apex angle=60, minimum size=12mm, inner sep=1pt, rotate=0},
  core_raw/.style={
    draw=black!50,
    rectangle,
    minimum width=20mm,
    minimum height=10mm,
    fill=BEmat,
    inner sep=1pt,
    path picture={
      \draw[black!20,very thin] (path picture bounding box.north west) -- (path picture bounding box.south east);
    }
  },
  core_mid/.style={draw=black!60, rectangle, minimum width=20mm, minimum height=10mm, fill=BEcoreA!18!BEunit!8, inner sep=1pt},
  core_nearunit/.style={draw=black!60, rectangle, minimum width=20mm, minimum height=10mm, fill=BEunit!14, inner sep=1pt},
  padfill/.style={draw=none, fill=BEpad},
  ancilla/.style={draw=black!50, fill=black!12, circle, minimum size=6pt},
  bundle/.style={line width=1.6pt, -Stealth, draw=black!75},
  thinleg/.style={-Stealth, thin, draw=black!75},
  dimleg/.style={-Stealth, thin, draw=black!75},
  small/.style={font=\footnotesize},
  core_unitary/.style={
    draw=BEunit!60!black,
    fill=BEunit!12,
    regular polygon,
    regular polygon sides=3,
    minimum size=14mm,
    inner sep=1pt,
    shape border rotate=45, 
  },
  sw_active/.style={-Stealth, line width=1pt, black},
  sw_exhaust/.style={-Stealth, line width=0.9pt, gray!50},
  sw_und/.style={line width=0.8pt, gray!60},
  bond/.style={thin, draw=black!75},
 circuitblock/.style={
  draw=black!65,
  rounded corners=1.2pt,
  rectangle,
  minimum width=16mm,
  minimum height=12mm,
  fill=BEgen!35!BEmat,
  inner sep=1pt,
},
wire/.style={
  line width=1.0pt,
  draw=black!80,
},
bondwire/.style={
  line width=1.3pt,
  draw=black!80,
},
sweepunitary/.style={
  unitary,
  minimum size=9mm,
},
sweeptensor/.style={
  generaltensor,
  minimum width=8mm,
},
}

\tikzset{
  dirtriangle/.style={
    draw=BEunit!60!black,
    fill=BEunit!10,
    isosceles triangle,
    isosceles triangle apex angle=60,
    minimum size=12mm,
    inner sep=1pt,
  },
  dirtriangle up/.style={dirtriangle, shape border rotate=90},
  dirtriangle right/.style={dirtriangle, shape border rotate=0},
  dirtriangle down/.style={dirtriangle, shape border rotate=270},
  dirtriangle left/.style={dirtriangle, shape border rotate=180},
}

\title{Tensor Networks as an Explicit Interface for Quantum Block-Encodings}
\author{Sebastian Issel}
\date{August 2026}

\begin{document}
\maketitle

\begin{abstract}
Tensor networks (TNs) give explicit classical descriptions of structured finite linear maps, while block-encodings (BEs) are the standard quantum access model for such maps.
We establish TNs as a universal data interface for quantum algorithms: any explicitly specified TN for an arbitrary finite linear map compiles directly to an explicit qubit BE, with no penalty on the selected-block round trip under faithful compression.

Along a chosen sweep, the compiler handles local non-unitarity by exact local dilations and aggregates the resulting post-selection conditions online using logarithmically many additional flag qubits.
The chosen sweep incurs three exact, sweep-dependent costs (the accumulated scale, the frontier memory, and the number of genuinely dilated local steps), which the main theorems are stated in terms of.

In the bounded-local explicit arithmetic model, this yields linear-time compilation to linear-size circuits with constant-size local gadgets, and hence a constant-factor size correspondence between bounded-local TNs and bounded-local BEs.

The same construction gives a selected-block round trip \(\mathrm{BE}\to\mathrm{TN}\to\mathrm{BE}\):
a BE canonically yields a TN for its selected block rather than for an arbitrary unitary extension, which can then be compressed or approximated classically before recompilation.
Faithful Schmidt-rank compression transfers monotonically (the scale cost and frontier memory cannot increase, with operator error bounded by the discarded weight), while arbitrary restructuring admits no general scale guarantee, a limitation we show is unavoidable.

As consequences and limits of this scale accounting, we characterize exact scale optimality, show that bridge-hourglass forests admit scale-optimal sweeps after exact recursive local bond compression, and prove that certifying unrestricted exact scale optimality is already hard for diagonal MPOs on a path unless \(\classP=\classNP\).
\end{abstract}

\section{Introduction}

Many structured operators used in quantum algorithms are available first as explicit classical tensor-network data rather than as hand-designed unitary circuits.
At the same time, block-encoding is the standard quantum access model for using such operators inside quantum algorithms.
This creates a basic interface problem:
given an explicit TN for a finite linear map, when can it be compiled directly into an explicit qubit BE?

This question is not limited to one-dimensional MPOs or to square operators.
Useful operator data are often rectangular, and reducing a general TN to one-dimensional form can destroy the native graph structure that made the representation compact in the first place.
States and effects, encoders and decoders, projections, and transfer maps between spaces of different dimension should be treated as native objects, not as artifacts of auxiliary square embeddings.

This paper gives such a compiler.
It processes an explicit TN site by site along a chosen sweep and compiles the represented finite linear map directly to an explicit qubit BE for arbitrary network geometry.
Local non-unitarity is handled by exact local dilations, and the resulting post-selection conditions can be aggregated online using only logarithmically many additional flag qubits.
For the compiled BE, the construction keeps exact bookkeeping of the resulting scale, frontier memory, and genuinely dilated local steps.
Figure~\ref{fig:workflow} sketches the compiler and the selected-block optimization workflow built on it.

In the bounded-local regime, where site degree and leg dimension are uniformly bounded, the compiler runs in linear time and produces linear-size circuits built from constant-size local gadgets.
Combined with the standard circuit-to-TN construction, this yields a constant-factor size correspondence between bounded-local TNs and bounded-local BEs in the explicit model.

The same construction also closes the round trip
\[
\mathrm{BE}\to\mathrm{TN}\to\mathrm{BE}.
\]
An explicit BE canonically determines a TN for its selected block, so classical TN manipulations act directly on the represented operator rather than on an arbitrary unitary extension.
This gives an operator-level optimization route:
one may start from any explicit BE, convert it to a TN for its selected block, perform exact or approximate TN compression and refactorization there, and then recompile the result to a new explicit BE.
In this way, locally unnecessary support, bond rank, or small tensor fragments can often be removed before recompilation.
Under faithful Schmidt-rank compression the improvement transfers provably and monotonically to the recompiled BE, while arbitrary restructuring admits no general guarantee, a limitation we show in Section~\ref{sec:structured} to be unavoidable.
More broadly, this identifies operator TNs as a concrete explicit-data interface for classical operator descriptions supplied to quantum algorithms.

Finally, we study exact scale optimization around this compiler interface.
We characterize exact scale optimality, identify a structured positive regime for bridge-hourglass forests, the class whose non-isometric residual concentrates at a single bridge site after exact recursive local bond compression, and show that certifying unrestricted exact scale optimality is already hard for diagonal MPOs on a path unless \(\classP=\classNP\).

All exact compilation statements are made in an explicit arithmetic model with arbitrary exact one- and two-qubit gates; finite-gate-set synthesis and bit-complexity overheads are outside the main theorem unless stated otherwise.

\begin{figure}[t]
\centering
\begin{tikzpicture}[
  x=1mm,y=1mm,
  every node/.style={inner sep=1pt},
  flow/.style={-Stealth, line width=0.5pt, draw=black!55},
  lab/.style={font=\scriptsize, text=black!70, align=center}
]

\begin{scope}[shift={(10,44)}]
  \node[generaltensor, minimum width=5.6mm] (t1) at (0,4) {};
  \node[generaltensor, double, double distance=0.5pt, minimum width=5.6mm] (t2) at (9,10) {};
  \node[generaltensor, minimum width=5.6mm] (t3) at (9,-2) {};
  \node[generaltensor, minimum width=5.6mm] (t4) at (20,4) {};

  \draw[sw_und] (t1) -- (t2);
  \draw[sw_und] (t1) -- (t3);
  \draw[sw_und] (t2) -- (t4);
  \draw[sw_und] (t3) -- (t4);

  \draw[sw_und] ($(t1)+(-5.5,0)$) -- (t1);
  \draw[sw_und] ($(t3)+(-4,-5)$) -- (t3);
  \draw[sw_und] (t2) -- ++(4,6);
  \draw[sw_und] (t4) -- ++(5.5,1.5);
  \draw[sw_und] (t4) -- ++(5.5,-3.5);

  \node[font=\scriptsize, anchor=west] at ($(t2)+(2.4,3)$) {\(\pi_j\)};
  \node[lab] at (10,-10) {TN};
\end{scope}

\draw[flow] (40,48) -- (78,48);

\begin{scope}[shift={(60,55)}]
  \node[circuitblock, minimum width=16mm, minimum height=12mm] (Q) at (0,0) {};
  \node[matrix,
        minimum width=7.5mm,
        minimum height=4.7mm,
        font=\tiny,
        anchor=north west] (A) at (Q.north west) {\(A_{\pi_j}\)};

  \coordinate (qLbase) at ($(Q.west)+(-3.2,0)$);
  \coordinate (qRbase) at ($(Q.east)+(3.2,0)$);

  \coordinate (qtopL) at (qLbase |- A.center);
  \coordinate (qtopR) at (qRbase |- A.center);
  \draw[wire] (qtopL) -- (A.west);
  \draw[wire] (A.east) -- (qtopR);

  \coordinate (qauxY)  at ($(Q.center)+(0,-2.3)$);
  \coordinate (qauxLW) at (Q.west |- qauxY);
  \coordinate (qauxRW) at (Q.east |- qauxY);
  \coordinate (qauxL)  at (qLbase |- qauxY);
  \coordinate (qauxR)  at (qRbase |- qauxY);

  \draw[wire] (qauxL) -- (qauxLW);
  \draw[wire] (qauxRW) -- (qauxR);

  \node[font=\scriptsize, anchor=east] at ($(qauxL)+(-0.6,0)$) {\(\ket{0}\)};
  \node[font=\scriptsize, anchor=west] at ($(qauxR)+(0.6,0)$) {\(\bra{0}\)};

  \node[font=\scriptsize] at ($(Q.center)+(2.0,-0.8)$) {\(Q_{\pi_j}\)};
\end{scope}

\begin{scope}[shift={(94,45)}]
  \draw[wire] (-7,8) -- (7,8);
  \draw[wire] (-7,3) -- (7,3);
  \draw[wire] (-7,-2) -- (7,-2);

  \node[font=\scriptsize, anchor=east] at (-8,8) {\(\ket{0}\)};
  \node[font=\scriptsize, anchor=west] at (8,8) {\(\bra{0}\)};

  \node[circuitblock, minimum width=4.8mm, minimum height=9mm] at (-3.0,5.5) {\(\scriptstyle U\)};
  \node[circuitblock, minimum width=4.8mm, minimum height=9mm] at ( 3.0,0.5) {\(\scriptstyle V\)};

  \node[lab] at (0,-10.5) {compiled BE};
\end{scope}

\begin{scope}[shift={(8,12)}]
  \draw[wire] (-7,8) -- (7,8);
  \draw[wire] (-7,3) -- (7,3);
  \draw[wire] (-7,-2) -- (7,-2);

  \node[font=\scriptsize, anchor=east] at (-8,8) {\(\ket{0}\)};
  \node[font=\scriptsize, anchor=west] at (8,8) {\(\bra{0}\)};

  \node[circuitblock, minimum width=4.8mm, minimum height=9mm] at (-3.0,5.5) {\(\scriptstyle U\)};
  \node[circuitblock, minimum width=4.8mm, minimum height=9mm] at ( 3.0,0.5) {\(\scriptstyle V\)};

  \node[lab] at (0,-10.5) {BE};
\end{scope}

\draw[flow] (18.5,15) -- (25.5,15);

\begin{scope}[shift={(44,10)}]
  \node[generaltensor, minimum width=5mm] (b1) at (-9,4) {};
  \node[generaltensor, minimum width=5mm] (b2) at (0,10) {};
  \node[generaltensor, minimum width=5mm] (b3) at (0,-2) {};
  \node[generaltensor, minimum width=5mm] (b4) at (9,4) {};

  \draw[sw_und] (b1) -- (b2);
  \draw[sw_und] (b1) -- (b3);
  \draw[sw_und] (b2) -- (b4);
  \draw[sw_und] (b3) -- (b4);

  \draw[sw_und] ($(b1)+(-4.5,0)$) -- (b1);
  \draw[sw_und] ($(b3)+(-3.3,-4.2)$) -- (b3);
  \draw[sw_und] (b2) -- ++(3.2,5.0);
  \draw[sw_und] (b4) -- ++(4.5,1.0);
  \draw[sw_und] (b4) -- ++(4.5,-3.2);

  \node[font=\scriptsize] at (0,4) {\(B\)};
\end{scope}

\draw[flow] (60,15) -- (66,15);
\node[lab] at (63,1.5) {compress};

\begin{scope}[shift={(79,10)}]
  \node[generaltensor, minimum width=4.6mm] (c1) at (-4,7) {};
  \node[generaltensor, minimum width=4.6mm] (c2) at ( 4,0) {};

  \draw[sw_und, line width=0.9pt] (c1) -- (c2);

  \draw[sw_und] ($(c1)+(-4.2,0)$) -- (c1);
  \draw[sw_und] ($(c1)+(-3.0,4.0)$) -- (c1);
  \draw[sw_und] (c2) -- ++(4.2,0);
  \draw[sw_und] (c2) -- ++(3.0,-4.0);

  \node[font=\scriptsize] at (2,5) {\(B'\)};
\end{scope}

\draw[flow] (89,15) -- (94,15);

\begin{scope}[shift={(102,12)}]
  \draw[wire] (-7,8) -- (7,8);
  \draw[wire] (-7,3) -- (7,3);
  \draw[wire] (-7,-2) -- (7,-2);

  \node[font=\scriptsize, anchor=east] at (-8,8) {\(\ket{0}\)};
  \node[font=\scriptsize, anchor=west] at (8,8) {\(\bra{0}\)};

  \node[circuitblock, minimum width=4.8mm, minimum height=9mm] at (-3.0,5.5) {\(\scriptstyle W\)};
  \node[circuitblock, minimum width=4.8mm, minimum height=9mm] at ( 3.0,0.5) {\(\scriptstyle X\)};

  \node[lab] at (0,-10.5) {optimized BE};
\end{scope}

\end{tikzpicture}
\caption{Compiler and selected-block optimization workflow.
Top: the TN is processed along a sweep; at the highlighted site \(\pi_j\), the local unfolded map \(A_{\pi_j}\) is embedded into a selected-block gadget \(Q_{\pi_j}\).
Composing these local steps yields a compiled BE for \(H(\mathcal T)\), together with the sweep-dependent quantities \(\Gamma(\pi)\), \(M(\pi)\), and \(D(\pi)\).
Bottom: starting from a BE, fixing the boundary conditions yields a TN for the selected block \(B\); this TN can be compressed, refactorized, or approximated classically to a new TN \(B'\), which is then recompiled to an optimized BE.}
\label{fig:workflow}
\end{figure}

The paper is organized as follows.
Section~\ref{sec:related} reviews related work.
Section~\ref{sec:results} defines the model and states the main results.
Section~\ref{sec:local} develops the local realization primitives.
Section~\ref{sec:global} proves the global compiler and the online flag aggregation bound.
Section~\ref{sec:complexity} derives the compiler resource bounds, transfer principles, and the bounded-local correspondence result.
Section~\ref{sec:structured} studies exact scale optimality, proves the bridge-hourglass positive result, and gives the hardness barrier for unrestricted exact optimization.
Section~\ref{sec:discussion} concludes.

\section{Related work}\label{sec:related}

BEs are a standard primitive in quantum algorithms for Hamiltonian simulation, linear algebra, and operator manipulation \cite{Low2019,Gilyen2019}.
Our focus here is different:
we study compilation from explicit classical operator descriptions.
Concretely, given an explicitly specified TN for an operator, when can it be turned systematically into an explicit BE?

At the local level, our realization primitives use the standard Halmos unitary dilation of a finite-dimensional contraction \cite{Halmos1950}.
At the network level, we also rely on the familiar correspondence between quantum circuits and TNs, as well as the view of tensor contraction as a postselected circuit computation \cite{Markov2008}.
For the structured tree preprocessing used later, we invoke standard rooted canonical/isometric forms for tree TNs \cite{Shi2006,Murg2010,Schollwoeck2011,Orus2014}.

TNs are a standard classical language for structured many-body states and operators, including matrix product states, matrix product operators, tree TNs, and circuit-derived networks \cite{Schollwoeck2011,Orus2014}.
For operator representations, MPO constructions for long-range Hamiltonians and related structured models are especially well developed \cite{Crosswhite2008,Pirvu2010}.
We use this literature as a source of explicit operator descriptions and optimization methods.

Variational quantum algorithms usually specify the ansatz directly as a parameterized circuit \cite{Cerezo2021}.
The present interface also allows a TN-first route: choose a parameterized operator TN with the desired graph connectivity, optimize its local tensors as explicit data, and compile the resulting map to a BE.

TN/circuit duality, postselected-circuit viewpoints on tensor contraction, and local unitary dilations are standard ingredients.
The point of the present paper is not those ingredients in isolation, but their organization into an explicit operator-level compiler interface.
What is compiled here is not a scalar contraction instance and not merely a square MPO, but an arbitrary explicit TN for a finite linear map, including native rectangular maps and arbitrary network geometry.
The compiler makes the resulting BE costs explicit through the chosen sweep, exposing exact accumulated scale, frontier memory, and dilation bookkeeping, and it adds an online logarithmic flag-reuse mechanism rather than leaving all primitive post-selection conditions live.
Combined with the standard circuit-to-TN construction, this also yields a bounded-local TN/BE round trip in which the intermediate TN represents the selected block itself, so classical TN manipulations can act directly at the operator level before recompilation.

The closest prior work in scope is the MPO-specific BE construction of Nibbi and Mendl \cite{Nibbi2024}.
Their setting is specialized to one-dimensional MPOs, and the construction is organized around the uniform MPO structure.
Restricting the present compiler to a one-dimensional path graph, padding the MPO bonds to a constant virtual dimension, and fixing a canonical left-to-right (or right-to-left) sweep specializes it to a per-site MPO dilation in the construction family of \cite{Nibbi2024}; the online flag aggregation of Section~\ref{sec:reuse-log} then improves the per-site ancilla accounting to logarithmic in the number of genuinely dilated steps.
Related but more restrictive MPO-to-circuit and TN-based optimization constructions also appear in \cite{Termanova2024,Akshay2024}, but not in the form of a general TN-to-BE compiler in the explicit classical input model considered here.

A concurrent and independent preprint of Dumitrescu \cite{Dumitrescu2026} develops an MPO block-encoding compiler that treats an MPO as a compressed linear-combination-of-unitaries program and builds conditional PREP and local SELECT stages from a parent MPO, with a numerical study of real-time Heisenberg-chain evolution.
That work shares the broad motivation of viewing TNs as compiler intermediate representations for block encoding, but it remains within the one-dimensional MPO/LCU setting.
By contrast, our construction applies to arbitrary explicitly specified TN geometries and native rectangular maps, does not reduce to an LCU over an explicit unitary list, and is organized around two structural ingredients:
a local per-site BE theorem for unfolded site operators, and a global sweep composition theorem that assembles these local gadgets according to the contraction pattern.
This separation is what makes the construction geometry-agnostic.
It also enables the bounded-local correspondence result, the exact sweep-dependent resource accounting, and the transfer of TN preprocessing methods into BE optimization.

Our bounded-local correspondence result places standard quantum circuits and bounded-local TNs into a common explicit bounded-local framework.
Viewing a bounded-local circuit as a TN is immediate.
To our knowledge, the converse direction, namely compiling bounded-local TN descriptions into bounded-local BE circuits with only constant-factor overhead, has not been isolated in this explicit bounded-local form in the existing TN or BE literature.

The online flag-aggregation gadget of Section~\ref{sec:global} aggregates streaming post-selection success conditions into a logarithmic number of reusable flag slots.
It is an exact, online streaming-AND construction for the compiler's success flags, and is distinct from the broader dirty-ancilla and conditionally-clean-ancilla reuse literature \cite{Khattar2024}, which targets general circuit constructions rather than the streaming aggregation of dilation success flags studied here.

\section{Model and main results}\label{sec:results}

We work throughout in an explicit classical input model for finite linear maps.
The main compiler problem takes as input an explicitly specified TN representing a finite linear map
\[
H:\mathcal H_{\mathrm{in}}\to\mathcal H_{\mathrm{out}}
\]
and produces an explicit BE for that same map.
Unless stated otherwise, all compilation and correspondence statements are arithmetic-complexity statements over \(\C\) on explicitly stored scalars.
Throughout, ``constructs'' means in the unit-cost exact arithmetic model over explicitly stored complex scalars.
Dense linear-algebra primitives such as SVDs, spectral norms, matrix square roots, and exact zero/support tests are charged by arithmetic-operation count, and output gates and scale factors may contain arbitrary exact complex entries.
No finite-gate-set synthesis or bit-complexity claim is made except in Proposition~\ref{prop:main-hardness}.

\subsection{Explicit model}

A \emph{block-encoding} (BE) in this paper may represent any finite linear map
\[
H:\mathcal H_{\mathrm{in}}\to\mathcal H_{\mathrm{out}},
\]
so square operators are only the special case
\[
\mathcal H_{\mathrm{in}}=\mathcal H_{\mathrm{out}}.
\]
A BE consists of:
\begin{itemize}
\item a qubit circuit \(U\) built from one- and two-qubit unitary gates,
\item a positive scale factor \(\alpha\),
\item designated input and output qubit registers, interpreted at the initial and final time boundaries, and
\item designated ancilla qubits together with their initialization and post-selection values.
\end{itemize}
The initialization and post-selection ancilla sets may overlap.

Let \(\mathcal K\) be the ambient qubit Hilbert space of \(U\).
The boundary data define canonical isometries
\[
R_{\mathrm{in}}:\mathcal H_{\mathrm{in}}\to\mathcal K,
\qquad
R_{\mathrm{out}}:\mathcal H_{\mathrm{out}}\to\mathcal K,
\]
and hence the selected block
\[
B:=R_{\mathrm{out}}^\dagger U R_{\mathrm{in}},
\qquad
H=\alpha B.
\]
Rectangular maps arise already at the local level: the sweep-induced unfolded site operators are generally rectangular.
The global BE simply inherits that native operator structure, rather than forcing an artificial square reformulation with extra ancillas and boundary conditions.

We measure BE size by
\[
\mathrm{size}_{\mathrm{BE}}
:=
\#\text{gates}
+
\#\text{qubits}
+
\#\text{boundary items}
+
1,
\]
where the boundary items are the input/output register declarations and the ancilla initialization/post-selection declarations, and the final \(1\) counts the scale factor.
Since each one- or two-qubit gate carries only constant-size matrix data, this agrees up to constants with counting the stored gate entries.
We work in an abstract qubit circuit model.
If a fixed native gate set is imposed afterwards, the usual exact or approximate synthesis overhead is additional.

We take \(\alpha>0\).
Zero operators are still allowed; in that case we use the canonical convention \(\alpha=1\) and selected block \(0\).
The degenerate zero-gauge \(\alpha=0\) is deferred to Appendix~\ref{app:zero}.

A \emph{tensor network} (TN) consists of site tensors together with incidence data.
For each site \(v\), one specifies a tensor \(T^{(v)}\) and an ordered list of its incident legs.
Each leg is either paired with exactly one leg of another site, forming an internal bond, or marked as a global input leg or a global output leg.
Equivalently, after adjoining a virtual source \(s\) for all input legs and a virtual sink \(r\) for all output legs, the combinatorial data are the sitewise adjacency lists in the resulting augmented graph.
Contracting all internal bonds yields the represented linear map, denoted
\[
H(\mathcal T).
\]

Figure~\ref{fig:TN} illustrates this convention.

\begin{figure}[ht]
\centering
\begin{tikzpicture}[x=1mm,y=1mm,every node/.style={font=\small,inner sep=1pt}]

\node[generaltensor, minimum width=8mm] (v1) at (0,12) {\(T^{(v_1)}\)};
\node[generaltensor, minimum width=8mm] (v2) at (22,22) {\(T^{(v_2)}\)};
\node[generaltensor, minimum width=8mm] (v3) at (24,2) {\(T^{(v_3)}\)};
\node[generaltensor, minimum width=8mm] (v4) at (46,12) {\(T^{(v_4)}\)};

\draw[sw_und] (v1) -- (v2);
\draw[sw_und] (v1) -- (v3);
\draw[sw_und] (v2) -- (v4);
\draw[sw_und] (v3) -- (v4);

\draw[sw_und] ($(v1)+(-12,0)$) -- (v1);
\node[font=\scriptsize, anchor=east] at ($(v1)+(-13,0)$) {\(i_1\)};

\draw[sw_und] ($(v3)+(-8,-10)$) -- (v3);
\node[font=\scriptsize, anchor=east] at ($(v3)+(-9,-10)$) {\(i_2\)};

\draw[sw_und] (v2) -- ++(8,10);
\node[font=\scriptsize, anchor=south] at ($(v2)+(9,10)$) {\(o_1\)};

\draw[sw_und] (v4) -- ++(12,4);
\node[font=\scriptsize, anchor=west] at ($(v4)+(13,4)$) {\(o_2\)};

\draw[sw_und] (v4) -- ++(12,-6);
\node[font=\scriptsize, anchor=west] at ($(v4)+(13,-6)$) {\(o_3\)};

\end{tikzpicture}
\caption{TN convention used in this paper.
Internal edges represent contractions.
Every free leg is labeled as a global input or output, so the full contraction defines an operator \(H(\mathcal T)\).}
\label{fig:TN}
\end{figure}

We use the usual tensor-data size
\[
\mathrm{size}(\mathcal T):=\sum_{v\in V}|T^{(v)}|,
\]
that is, the total number of stored tensor entries.
The incidence lists and input/output labels are also part of the input, but they play only a bookkeeping role in the statements below.
If desired, they may be added additively to the size measure without changing any later polynomial-overhead or bounded-local result.

All leg Hilbert spaces are assumed finite-dimensional and nonzero-dimensional.
The empty tensor product is \(\C\).
If \(V=\emptyset\), the TN represents the scalar identity map \(\C\to\C\), i.e. the scalar \(1\).
All nontrivial operator-valued networks considered below have \(L\ge 1\).

For simplicity, the main text uses the qubit-register convention in which each leg Hilbert space is embedded by zero extension into a power-of-two qubit space.
The general-dimension variant is deferred to Appendix~\ref{app:padding}.

\subsection{Bounded-local regime}

The bounded-local regime is the setting most relevant for comparison with standard quantum circuits.

On the TN side, bounded-local means for all sites a uniformly bounded site degree, and uniformly bounded leg dimension.

On the BE side, bounded-local means a qubit circuit built from arbitrary one- and two-qubit unitary gates, with boundary data as in the BE model above.

In a bounded-local TN, each site tensor has constant size, so
\[
\mathrm{size}(\mathcal T)=\Theta(L).
\]
Thus asymptotic complexity in TN size is equivalent to asymptotic complexity in the number of sites.

We restrict attention to finite operator data given classically and to bounded-local quantum procedures; black-box and QRAM access models are outside scope.

\subsection{Sweeps and sweep-dependent quantities}

The compiler processes the TN site by site along a chosen sweep
\[
\pi=(v_1,\dots,v_L),
\]
that is, an ordering of the vertex set \(V\).

That sweep determines:
\begin{itemize}
\item how each site tensor is unfolded into a local operator,
\item the sequence in which local gadgets are composed,
\item the frontier carried across the sweep cut, and
\item which local steps require genuine one-flag dilations.
\end{itemize}
Thus the compiler works on the native TN graph rather than on a forced one-dimensional reduction.
The chosen sweep makes the layout cost explicit through the frontier it carries, rather than hiding that cost inside a prior MPO conversion.

Three sweep-dependent quantities will be tracked for the compiled BE:
\begin{itemize}
\item \(\Gamma(\pi)\), the global scale accumulated along the sweep;
\item \(M(\pi)\), the memory carried by the sweep frontier; and
\item \(D(\pi)\), the number of genuinely dilated local steps under the default local realization policy of Section~\ref{sec:local}.
\end{itemize}

These symbols are introduced here only to name the quantities that appear in the main theorems.
Their precise definitions are given later in Section~\ref{sec:global}.

\subsection{Main results}

The novelty is therefore not the separate existence of local dilations or of circuit/TN translations, both of which are standard, but the resulting explicit compiler interface for operator TN data.
In particular, the construction works directly on the native TN geometry, treats rectangular local and global maps without forcing a prior square reformulation, and isolates the sweep-dependent resources that govern the compiled BE.
This is what later enables the online flag aggregation theorem, the bounded-local correspondence, and the selected-block recompilation viewpoint.

We begin with the direct compiler theorem.
\begin{theorem}[Direct TN-to-BE compiler]\label{thm:main-compiler}
Let \(\mathcal T\) be an explicitly specified TN on vertex set \(V\), and let
\[
\pi=(v_1,\dots,v_L)
\]
be any sweep order of \(V\).
Let \(\beta_t\) denote the sweep-induced local spectral scales defined in Section~\ref{sec:global}.

If \(\beta_t=0\) for some \(t\), then \(H(\mathcal T)=0\), and the compiler may immediately return the canonical zero BE with scale \(1\) and selected block \(0\).

Otherwise, all \(\beta_t>0\), and one can construct, using a number of arithmetic operations polynomial in the explicit input length of \(\mathcal T\), an explicit BE of polynomial size with scale
\[
\Gamma(\pi)=\prod_{t=1}^{L}\beta_t
\]
whose represented operator is \(H(\mathcal T)\).
Equivalently, its selected block \(\widehat H_\pi\) satisfies
\[
H(\mathcal T)=\Gamma(\pi)\,\widehat H_\pi.
\]
Concrete compile-time and circuit-size bounds are stated later in Theorem~\ref{thm:compile} and Theorem~\ref{thm:bounded-linear}.

The construction also exposes three exact sweep-dependent resource quantities:
the accumulated scale \(\Gamma(\pi)\), the frontier memory \(M(\pi)\), and the number \(D(\pi)\) of genuinely dilated local steps.
\end{theorem}

\begin{theorem}[Online logarithmic flag aggregation]\label{thm:main-flags}
In the compiled BE of Theorem~\ref{thm:main-compiler}, the primitive success flags introduced by the \(D(\pi)\) genuinely dilated local steps can be aggregated online using only
\[
\bigO{\log(D(\pi)+1)}
\]
additional reusable dirty-flag qubit slots.

More precisely, the total memory is
\[
M(\pi)+\bigO{\log(D(\pi)+1)},
\]
where the logarithmic term is exactly the online dirty-flag slot pool used to hold primitive or merged success conditions.
The number of online flag-merge gadgets is at most \(D(\pi)-1\) when \(D(\pi)\ge 1\), and \(0\) when \(D(\pi)=0\).
\end{theorem}

As a consequence, the bounded-local regime admits a sharp comparison with standard bounded-local BE circuits.

\begin{theorem}[Bounded-local TN/BE correspondence for arbitrary finite linear maps]\label{thm:main-correspondence}
In the explicit classical input model, with fixed locality constants and arbitrary one- and two-qubit gates on the BE side, bounded-local TNs and bounded-local BEs correspond up to constant-factor overhead in circuit size.
This is the same statement as Theorem~\ref{thm:correspondence}, restated here among the main results and proved in Section~\ref{sec:complexity}.

More explicitly, after fixing the locality bounds and the constant-size local gate conventions, there are constants independent of \(T\) such that:
\begin{itemize}
\item every bounded-local TN of size \(T\) compiles to a bounded-local BE of size at most a constant multiple of \(T\), and
\item every bounded-local BE of size \(T\) canonically yields a bounded-local TN of size at most a constant multiple of \(T\).
\end{itemize}
\end{theorem}

Thus, within this model, bounded-local TN representability and bounded-local BE realizability coincide up to constant-factor overhead in circuit size (the subnormalization \(\alpha\) is tracked separately throughout).
Corollary~\ref{cor:round-trip} later gives the corresponding selected-block round trip
\[
\mathrm{BE}\to\mathrm{TN}\to\mathrm{BE},
\]
whose practical point is that the intermediate TN represents the selected block \(B=\alpha^{-1}H\) itself rather than an arbitrary unitary dilation.

We next turn to exact resource improvement and its limits.

\begin{proposition}[Scale-optimality criterion]\label{prop:main-scale}
Fix a TN \(\mathcal T\) and a sweep \(\pi\), and assume \(H(\mathcal T)\neq 0\).
Let \(\widehat H_\pi\) denote the selected block of the compiled BE, so that
\[
H(\mathcal T)=\Gamma(\pi)\,\widehat H_\pi.
\]
Then
\[
\Gamma(\pi)=\|H(\mathcal T)\|_2
\]
if and only if
\[
\|\widehat H_\pi\|_2=1.
\]
\end{proposition}

This is the convention used below for scale-optimality.

\begin{theorem}[Bridge-hourglass forests admit scale-optimal sweeps]\label{thm:main-hourglass}
Every bridge-hourglass forest admits a scale-optimal sweep after exact preprocessing using a number of arithmetic operations polynomial in the explicit input size, via recursive local bond compression.
\end{theorem}

\begin{proposition}[Exact scale-optimal preprocessing is hard already for diagonal MPOs]\label{prop:main-hardness}
Unless \(\classP=\classNP\), there is no unrestricted exact preprocessing procedure running in time polynomial in the total bit length of a binary-encoded diagonal MPO with integer entries on a path that always returns an equivalent TN representation, a sweep \(\pi\), and an exactly encoded scale value \(g\), comparable to integers in polynomial time, such that
\[
g=\Gamma(\pi)=\|H(\mathcal T)\|_2.
\]
\end{proposition}

\subsection{Roadmap to the proofs}

Section~\ref{sec:local} develops the local realization primitives.
It unfolds a site tensor into a local operator, normalizes it, embeds it into a square contraction, and realizes that contraction as a selected block.

Section~\ref{sec:global} assembles these local gadgets along a sweep, proves the selected-block correctness theorem, and proves the online logarithmic flag-aggregation bound.

Section~\ref{sec:complexity} derives the compiler resource bounds, analyzes sweep dependence and approximation transfer, proves the bounded-local correspondence theorem, and gives the selected-block round-trip consequences.

Section~\ref{sec:structured} proves Proposition~\ref{prop:main-scale}, Theorem~\ref{thm:main-hourglass}, and Proposition~\ref{prop:main-hardness}.

\section{Local realization primitives}\label{sec:local}

This section isolates the local ingredients used by the sweep compiler.
At a single site, the compiler:
\begin{enumerate}
\item unfolds the site tensor into a local operator,
\item normalizes that operator by its local spectral scale,
\item embeds the normalized map into a square contraction,
\item chooses a local unitary with that selected block, and
\item if a genuine dilation is used, manages the resulting primitive success flag.
\end{enumerate}

\subsection{Unfolded site operators}

For sitewise compilation, each site tensor is reshaped into an operator.
Figure~\ref{fig:unfold} illustrates the unfolding convention.

\begin{figure}[ht]
\centering
\begin{tikzpicture}[every node/.style={font=\small,inner sep=1pt}]
  \def\xsep{30mm}
  \def\descshift{20mm}

  \node[generaltensor] (Ta) at (0,0) {\(T^{(v)}\)};
  \foreach \ang in {20,60,100,140,200,240,280,320} {
    \draw (Ta) -- ($(Ta.center)+(\ang:15mm)$);
  }
  \node at ($(Ta)+(-10:15mm)$) {\(\vdots\)};
  \node at ($(Ta)-(-10:15mm)$) {\(\vdots\)};
  \node at ($(Ta)-(0,\descshift)$) {\textbf{(a)}};

  \node[directedtensor] (Tb) at ($(Ta.east)+(\xsep,0)$) {\(T^{(v)}\)};
  \node (Bbot) at ($(Tb)-(0,12mm)$) {\(\cdots\)};
  \node (Btop) at ($(Tb)+(0,12mm)$) {\(\cdots\)};
  \foreach \i in {-2,-1,1,2} {
    \draw[dimleg] ($(Bbot)+(0:\i*5mm)$) -- (Tb);
    \draw[dimleg] (Tb) -- ($(Btop)+(0:\i*5mm)$);
  }
  \node at ($(Tb)-(0,\descshift)$) {\textbf{(b)}};

  \node[matrix] (Tc) at ($(Tb.east)+(\xsep,0)$) {\(A^{(v)}\)};
  \draw[bundle] ($(Tc.south)+(0,-9mm)$) -- (Tc.south) node[midway,right]{\(n\)};
  \draw[bundle] (Tc.north) -- ++(0,9mm) node[midway,right]{\(m\)};
  \node at ($(Tc)-(0,\descshift)$) {\textbf{(c)}};
\end{tikzpicture}
\caption{Unfolding a site tensor into a matrix.
(a) A site tensor with incident legs.
(b) An unfolding convention assigns ordered input and output leg lists.
(c) Reshaping yields the unfolded site operator \(A^{(v)}\).}
\label{fig:unfold}
\end{figure}

Fix a site \(v\in V\).
An \emph{unfolding convention} at \(v\) is an ordered pair
\[
(I_v,O_v)
:=
\left((i_1,\dots,i_r),(o_1,\dots,o_s)\right)
\]
of disjoint leg lists whose union is the full incident leg set of \(v\).
This determines the spaces
\[
\mathcal H_{\mathrm{in}}^{(v)}
:=
\bigotimes_{j=1}^{r}\mathcal H_{i_j},
\qquad
\mathcal H_{\mathrm{out}}^{(v)}
:=
\bigotimes_{j=1}^{s}\mathcal H_{o_j},
\]
and hence an unfolded site operator
\[
A^{(v)}:\mathcal H_{\mathrm{in}}^{(v)}\to \mathcal H_{\mathrm{out}}^{(v)},
\qquad
A^{(v)}\in\C^{m_v\times n_v},
\]
with
\[
m_v:=\prod_{j=1}^{s}\dim(o_j),
\qquad
n_v:=\prod_{j=1}^{r}\dim(i_j).
\]

Equivalently, \(A^{(v)}\) is obtained by permuting the tensor indices of \(T^{(v)}\) so that the output legs appear in the order \(O_v\) and the input legs in the order \(I_v\), and then reshaping into a matrix with row dimension \(m_v\) and column dimension \(n_v\).

In the sweep compiler, the unfolding convention is induced by the chosen sweep:
bond legs attached to already processed sites are treated as inputs, bond legs attached to unprocessed sites as outputs, and physical legs retain their prescribed global input/output role.
Thus the unfolded operator generally depends on the sweep.
Reordering legs within \(I_v\) or within \(O_v\) changes only the tensor-product identification, whereas moving a leg between \(I_v\) and \(O_v\) changes the unfolded operator itself.

Define the local spectral scale
\[
\beta_v:=\|A^{(v)}\|_2.
\]
In the nondegenerate case \(\beta_v>0\), define the normalized local contraction
\[
C_v:=\beta_v^{-1}A^{(v)}.
\]
If \(\beta_v=0\), then certainly \(H(\mathcal T)=0\), and the compiler may immediately return the canonical zero BE of Appendix~\ref{app:zero}.

\subsection{Square embedding of the normalized local map}

To realize the normalized local map as a selected block of a unitary, we embed it into a square contraction.

Under the qubit convention of the main text, the unfolded dimensions have the form
\[
m_v=2^{a_v},
\qquad
n_v=2^{b_v}
\]
for integers \(a_v,b_v\ge 0\).
Hence
\[
\max(m_v,n_v)=2^{\max(a_v,b_v)}.
\]

Let \(\widetilde{\mathcal H}^{(v)}\) be a padded site register of dimension \(\max(m_v,n_v)\).
Choose canonical isometric embeddings
\[
J_{\mathrm{in}}:\mathcal H_{\mathrm{in}}^{(v)}\hookrightarrow \widetilde{\mathcal H}^{(v)},
\qquad
J_{\mathrm{out}}:\mathcal H_{\mathrm{out}}^{(v)}\hookrightarrow \widetilde{\mathcal H}^{(v)},
\]
and define the square embedded contraction
\[
\widetilde C_v:=J_{\mathrm{out}}\,C_v\,J_{\mathrm{in}}^\dagger.
\]
Equivalently, \(\widetilde C_v\) is obtained from \(C_v\) by zero-padding to a square matrix.
All local realizations used below are chosen so that their selected block on the padded site register is exactly \(\widetilde C_v\), not merely \(C_v\) on the abstract unpadded input/output spaces.

\begin{theorem}[Universal one-flag dilation primitive]\label{thm:local-be}
Let
\[
\widetilde C:\widetilde{\mathcal H}\to\widetilde{\mathcal H}
\]
be a contraction on a finite-dimensional Hilbert space.
Then there exists a unitary
\[
Q:\mathbb C^2\otimes\widetilde{\mathcal H}\to\mathbb C^2\otimes\widetilde{\mathcal H}
\]
such that
\[
(\bra{0}\otimes I_{\widetilde{\mathcal H}})\,Q\,(\ket{0}\otimes I_{\widetilde{\mathcal H}})
=
\widetilde C.
\]
One explicit choice is the Halmos dilation
\[
Q=
\begin{pmatrix}
\widetilde C & \sqrt{I-\widetilde C\widetilde C^\dagger}\\[1mm]
\sqrt{I-\widetilde C^\dagger \widetilde C} & -\widetilde C^\dagger
\end{pmatrix}.
\]
\end{theorem}

\begin{proof}
Since \(\widetilde C\) is a contraction, the displayed block matrix is unitary.
Its selected flag-\(\ket{0}\) block is exactly \(\widetilde C\).
\end{proof}

Applied to a site \(v\), Theorem~\ref{thm:local-be} gives a universal default local unitary realizing the normalized padded local map
\[
\widetilde C_v=J_{\mathrm{out}}\,C_v\,J_{\mathrm{in}}^\dagger
\]
as a selected block.

The one-flag realization is always available, but not always necessary.
Some normalized padded local maps already admit an unflagged unitary extension.

\begin{remark}[Unflagged padding-compatible extensions]\label{rem:unflagged-local}
Let
\[
C:\mathcal H_{\mathrm{in}}\to\mathcal H_{\mathrm{out}}
\]
be a normalized local map.
If
\[
C^\dagger C=I_{\mathcal H_{\mathrm{in}}},
\]
then \(C\) is an isometry.
In that case
\[
\widetilde C=J_{\mathrm{out}}\,C\,J_{\mathrm{in}}^\dagger
\]
is a partial isometry on the padded site register \(\widetilde{\mathcal H}\), and in finite dimension it extends to a unitary on \(\widetilde{\mathcal H}\).
We use such an extension in the unflagged case, so the selected block on the padded site register is exactly \(\widetilde C\).

A coisometry \(CC^\dagger=I_{\mathcal H_{\mathrm{out}}}\) also has local scale \(1\), but it is dimension-reducing.
Under the default policy below, it is still treated by the one-flag selected-block primitive unless it is also an isometry.
\end{remark}

We use the following default exact local realization policy for the resource counts below:
if the normalized local map \(C_v\) is an isometry, use an unflagged unitary extension;
otherwise use the one-flag realization primitive from Theorem~\ref{thm:local-be}.

Once a local unitary \(Q\) with the required selected block has been chosen, the compiler synthesizes it on the padded local register into one- and two-qubit gates, so the output is an explicit qubit circuit in the sense of Section~\ref{sec:results}.
For a dense \(q\)-qubit local unitary, standard QSD/CSD-type synthesis uses \(\bigO{4^q}\) one- and two-qubit gates \cite{Shende2004}.
A standard dense realization therefore contributes polynomial arithmetic overhead in the local dimension.
In the bounded-local regime \(q=\bigO{1}\), this is only a constant-factor local overhead.

For explicit preprocessing, one convenient local route computes a singular-value decomposition
\[
A^{(v)}=USV^\dagger.
\]
In this representation, only the singular-value core \(S\) contributes nontrivial attenuation.
The surrounding factors \(U\) and \(V^\dagger\) are unitary and do not reduce the selected branch.

If the normalized map is an isometry, equivalently \(C_v^\dagger C_v=I\), the local step is completed as an unflagged unitary extension.
Otherwise one uses a genuine dilation on the singular-value core.
Figure~\ref{fig:dilate} illustrates this SVD-based realization.
Whenever an exact support restriction is used, zero singular sectors are simply discarded.

\begin{figure}[ht]
\centering
\begin{tikzpicture}[every node/.style={font=\small,inner sep=1pt}]
  \def\xsep{40mm}
  \def\yoff{0mm}
  \def\descshift{45mm}

  \begin{scope}[xshift=0mm,yshift=\yoff]
    \node[core_raw] (Sdiag) at (0, 0) {\(S\)};
    \node[unitary, above=7mm of Sdiag] (U) {\(U\)};
    \node[unitary,below=7mm of Sdiag] (Vh) {\(V^\dag\)};
    \draw[bundle] ($(Vh.south)-(0,7mm)$) node[above right]{\(n\)} -- (Vh.south);
    \draw[bundle] (Vh.north) node[above right]{\(n\)} -- (Sdiag.south);
    \draw[bundle] (Sdiag.north) node[above right]{\(m\)} -- (U.south);
    \draw[bundle] (U.north) node[above right]{\(m\)} -- ($(U.north)+(0,7mm)$);
    \node[small, yshift=-\descshift] {(a)\(A=U\,S\,V^\dag\)};
  \end{scope}

  \begin{scope}[xshift=\xsep,yshift=\yoff]
    \node[square_matrix, rectangle, minimum width=25mm, minimum height=25mm] (Souter_e) {};
    \node[core_raw, anchor=north west] (Sinner_e) at (Souter_e.north west) {\(S\)};

    \node[unitary,above=7mm of Souter_e.north] (U_e) {\(U\)};
    \node[unitary,below=7mm of Souter_e.south] (Vh_e) {\(V^\dag\)};

    \draw[bundle] (Vh_e.south) ++(0,-8mm) node[above right]{\(n\)} -- (Vh_e.south);            
    \draw[bundle] (Vh_e.north) node[above right]{\(n\)} -- (Souter_e.south);                         
    \draw[bundle] (Souter_e.north) node[above right]{\(m\)} -- (U_e.south);
    \draw[bundle] (U_e.north) node[above right]{\(m\)} -- ++(0,8mm);              

    \coordinate (ExtraP_e) at ($(Souter_e.south east)+(6mm,-12mm)$);
    \draw[-Stealth] (ExtraP_e) -| (Souter_e.south east) node[near end,below right]{\(p\)};
    \coordinate (DropEnd_e) at ($(Souter_e.north west)+(-6mm,12mm)$);
    \draw[-Stealth] (Sinner_e.north west) |- (DropEnd_e) node[at start, above left]{\(q\)};

    \node[small,yshift=-\descshift] {(b) square + padding};
  \end{scope}

 \begin{scope}[xshift=2*\xsep,yshift=\yoff]
    \node[core_unitary, minimum size=20mm] (Core) {\(\mathcal{C}\)}; 

    \node[unitary, above=9mm of Core] (U_f) {\(U\)};
    \node[unitary, below=9mm of Core] (Vh_f) {\(V^\dag\)};
    \draw[bundle] (Vh_f.south) ++(0,-9mm) node[above right]{\(n\)} -- (Vh_f.south);
    \draw[bundle] (Vh_f.north) node[above right]{\(n\)} -- (Core.south);
    \draw[bundle] (Core.north) node[above right]{\(m\)} -- (U_f.south);
    \draw[bundle] (U_f.north) node[above right]{\(m\)} -- ++(0,8mm); 

    \draw[thinleg] ($(Core.east)+(8mm,0)$) -- (Core.east) node[midway,above]{\(c\)};
    \draw[thinleg] (Core.west) -- ++(-8mm,0) node[midway,above]{\(c\)};

    \draw[thinleg] ($(Core.south east)+(12mm,-8mm)$) -| ($(Core.south east)$) node[near end, below right]{\(p\)};
    \draw[thinleg] (Core.north west) |- ($(Core.west)+(-8mm,12mm)$) node[above left]{\(q\)};

    \node[small,yshift=-\descshift] {(c) Core unitary};
  \end{scope}
\end{tikzpicture}
\caption{SVD-based local realization.
The unfolded site operator is decomposed as \(A^{(v)}=USV^\dagger\).
The unitary factors \(U\) and \(V^\dagger\) align the action with the singular directions and do not introduce additional attenuation on the selected branch.
Only the singular-value core requires normalization-dependent treatment; \(p\) denotes padding, \(q\) discarded dimensions, and \(c\) the dilation flag channel.}
\label{fig:dilate}
\end{figure}

\subsection{Primitive flags, certification, and binary merging}

A genuine one-flag dilation introduces a primitive success flag.
When several such local steps are composed, it is useful to aggregate their success conditions online rather than keeping all primitive flags live until the end.

\begin{definition}[Dirty flag and certification]
Let \(S\) be the current set of explicitly selected flags, and let \(\ket{\Psi}\) be the current state.
A flag qubit \(c\) is \emph{certified zero by \(S\)} if
\[
(\bra{0}_S\otimes \bra{0}_c\otimes I)\ket{\Psi}
=
(\bra{0}_S\otimes I)\ket{\Psi}.
\]
Equivalently, once the flags in \(S\) are selected to \(0\), imposing the additional condition \(c=0\) does not change the selected branch.
A flag is \emph{dirty} if it is not yet certified zero by the current selected set and must therefore still be tracked explicitly.
\end{definition}

The basic aggregation primitive merges two dirty flags into one fresh flag.

\begin{lemma}[Binary flag-merge gadget]\label{lem:merge-gadget}
Let \(a\), \(b\), and \(c\) be flag qubits, with \(c\) initialized in \(\ket{0}\).
Apply \(X\) to \(c\), and then apply a doubly zero-controlled \(X\) from \(a\) and \(b\) onto \(c\).
Equivalently, apply the map
\[
\ket{a}\ket{b}\ket{c}
\mapsto
\ket{a}\ket{b}\ket{c\oplus a\oplus b\oplus ab}.
\]
Denote the resulting unitary by \(G_{ab\to c}\).
Then
\[
(I_{ab}\otimes \bra{0}_c)\,G_{ab\to c}\,(I_{ab}\otimes \ket{0}_c)
=
\ket{00}\!\bra{00}_{ab}.
\]
Hence post-selecting the flag \(c=0\) is exactly equivalent to post-selecting \(a=b=0\).
In particular, after the selected set is updated by replacing the conditions \(a=b=0\) with the single condition \(c=0\), the old flags \(a\) and \(b\) are certified zero and may be used as free slots.
\end{lemma}

\begin{proof}
The basis truth table is immediate:
the output flag is \(0\) exactly when \(a=b=0\), and it is \(1\) otherwise.
Thus selecting \(c=0\) projects exactly onto the \(ab=00\) sector.
This is precisely the certification condition for \(a\) and \(b\) relative to the updated selected set.
\end{proof}

\begin{lemma}[Certified-zero reuse]\label{lem:certified-reuse}
Let
\[
K:\mathcal H_{\mathrm{in}}\to \mathcal H_S\otimes\mathcal H_z\otimes\mathcal H_R
\]
be the current compiled prefix map, where \(S\) is the current selected flag set.
Assume that \(z\) is certified zero by \(S\), i.e.
\[
(\bra{0}_S\otimes I_{zR})K
=
(\bra{0}_S\otimes \bra{0}_z\otimes I_R)K.
\]
Then for every later unitary or isometry
\[
V:\mathcal H_z\otimes\mathcal H_R\to\mathcal H_{R'}
\]
and every later selected set \(T\) introduced inside \(V\),
\[
(\bra{0}_T\otimes \bra{0}_S\otimes I)\,V\,K
=
(\bra{0}_T\otimes \bra{0}_S\otimes I)\,V\,(\ket{0}_z\otimes I_R)\,(\bra{0}_z\otimes I_R)K.
\]
In particular, on all later selected branches, the register \(z\) may be replaced by a fresh ancilla initialized in \(\ket{0}\), so its physical slot may be reused immediately.
\end{lemma}

\begin{proof}
Insert
\[
I_z=\ket{0}\!\bra{0}_z+\left(I_z-\ket{0}\!\bra{0}_z\right)
\]
before \(K\).
By the certification hypothesis, the second term vanishes after applying \(\bra{0}_S\).
The first term is exactly the claimed fresh-\(\ket{0}\) replacement.
\end{proof}

\section{Global compiler and online flag aggregation}\label{sec:global}

Fix a TN \(\mathcal T\) with \(L\) sites and a sweep order
\[
\pi=(v_1,\dots,v_L).
\]
The sweep determines the local unfolding at each site, the sequence of partial contractions, the frontier registers, and the placement of any primitive dilation flags.

\subsection{Sweep-induced frontier and local steps}

To define the sweep semantics, we will pass to the augmented graph obtained by adjoining a virtual source \(s\) and sink \(r\):
physical input legs become edges \((s,v)\), physical output legs become edges \((v,r)\), and internal bond edges remain unchanged.

For \(t=0,1,\dots,L\), let
\[
P_t:=\{v_1,\dots,v_t\},
\qquad
P_0:=\emptyset.
\]
Let \(F_t\) be the set of augmented edges crossing the cut
\[
P_t\cup\{s\}
\quad\text{and}\quad
(V\setminus P_t)\cup\{r\}.
\]
Equivalently, \(F_t\) consists of:
\begin{itemize}
\item bond legs connecting processed to unprocessed sites,
\item physical input legs incident to unprocessed sites, and
\item physical output legs incident to processed sites.
\end{itemize}

Thus \(F_t\) is exactly the interface across the sweep cut.
Define the sweep memory
\[
M(\pi):=\max_{0\le t\le L}\sum_{e\in F_t} q(e),
\]
where \(q(e)\) is the qubit count of leg \(e\).
This is the memory notion used throughout the paper.

Write
\[
\mathcal H(F_t):=\bigotimes_{e\in F_t}\mathcal H_e
\]
for the frontier Hilbert space at cut \(t\), with any fixed reference ordering of the frontier legs.

Figure~\ref{fig:sweep} shows one frontier update.

\begin{figure}[t]
\centering
\begin{tikzpicture}[
  x=1mm,y=1mm,
  every node/.style={font=\small,inner sep=1.5pt}
]

\def\panelsep{66}      
\def\labely{25}        
\def\arrowraise{1}     

\newcommand{\SweepPanel}[4]{%
  \begin{scope}[shift={#2}]
    \coordinate (#1O) at (0,0);

    \node at ($(#1O)+(0,\labely)$) {\textbf{#3}};

    \node[sweepunitary]      (#1b1) at ($(#1O)+(-18,  0)$) {};
    \node[sweepunitary]      (#1b2) at ($(#1O)+(-10, 10)$) {};
    \node[sweepunitary]      (#1b3) at ($(#1O)+(-10,-10)$) {};
    \node[sweepunitary]      (#1b4) at ($(#1O)+(-2,   6)$) {};

    \ifnum#4=0
      \node[sweeptensor, double] (#1hub) at ($(#1O)+(10,4)$) {};
    \else
      \node[sweepunitary]               (#1hub) at ($(#1O)+(10,4)$) {};
    \fi

    \node[sweeptensor] (#1r1) at ($(#1O)+(24, 10)$) {};
    \node[sweeptensor] (#1r2) at ($(#1O)+(30,  2)$) {};
    \node[sweeptensor] (#1r3) at ($(#1O)+(24, -8)$) {};

    \draw[sw_exhaust] (#1b1) to[bend left=12] (#1b2);
    \draw[sw_exhaust] (#1b2) to[bend left=10] (#1b4);
    \draw[sw_exhaust] (#1b3) to[bend left=10] (#1b1);

    \ifnum#4=0
      \draw[sw_und] (#1hub) -- (#1r1);
      \draw[sw_und] (#1hub) -- (#1r2);
      \draw[sw_und] (#1hub) -- (#1r3);
    \fi
    \draw[sw_und] (#1r1) -- (#1r2);
    \draw[sw_und] (#1r2) -- (#1r3);

    \ifnum#4=0
      \draw[sw_active] (#1b1) to[bend right=5]  (#1hub);
      \draw[sw_active] (#1b2) to[bend left=26]  (#1hub);
      \draw[sw_active] (#1b3) to[bend left=8]   (#1hub);
      \draw[sw_active] (#1b4) to[bend left=8]   (#1hub);
      \draw[sw_active] (#1b3) to[bend left=4]   (#1r3);
    \else
      \draw[sw_exhaust] (#1b1) to[bend right=5] (#1hub);
      \draw[sw_exhaust] (#1b2) to[bend left=26] (#1hub);
      \draw[sw_exhaust] (#1b3) to[bend left=8]  (#1hub);
      \draw[sw_exhaust] (#1b4) to[bend left=8]  (#1hub);

      \draw[sw_active] (#1hub) to[bend left=10]  (#1r1);
      \draw[sw_active] (#1hub) --                (#1r2);
      \draw[sw_active] (#1hub) to[bend right=10] (#1r3);
      \draw[sw_active] (#1b3)  to[bend left=4]   (#1r3);
    \fi

    \coordinate (#1west) at ($(#1O)+(-24,0)$);
    \coordinate (#1east) at ($(#1O)+(36,0)$);
  \end{scope}%
}

\SweepPanel{L}{(0,0)}{(a) before}{0}
\SweepPanel{R}{(\panelsep,0)}{(b) after}{1}

\node at ($(Least)!0.5!(Rwest)$) {\(\mapsto\)};

\end{tikzpicture}
\caption{Single sweep update.
(a) Before processing the next site, the frontier separates processed from unprocessed tensors.
(b) After processing that site, incoming frontier legs are retired and outgoing legs become the new frontier.
Active edges represent logical register labels.}
\label{fig:sweep}
\end{figure}

\begin{proposition}[Weighted cutwidth interpretation]\label{prop:cutwidth}
Let \(G_{\mathrm{aug}}\) be the augmented graph obtained from the TN by adjoining the virtual source \(s\) and sink \(r\), and let the edge weight of an augmented edge \(e\) be its qubit count \(q(e)\).
For the linear layout
\[
s,v_1,\dots,v_L,r
\]
induced by the sweep \(\pi\), the sweep memory satisfies
\[
M(\pi)
=
\max_{0\le t\le L}
\sum_{e\in \delta\!\left(P_t\cup\{s\}\right)} q(e),
\]
where \(\delta\!\left(P_t\cup\{s\}\right)\) denotes the augmented edges crossing the cut
\[
P_t\cup\{s\}
\quad\text{and}\quad
(V\setminus P_t)\cup\{r\}.
\]
Equivalently, \(M(\pi)\) is exactly the weighted cutwidth of the augmented graph under the sweep order.
\end{proposition}

\begin{proof}
By definition, \(F_t\) is precisely the set of augmented edges crossing that cut.
Therefore
\[
\sum_{e\in F_t} q(e)
=
\sum_{e\in \delta\!\left(P_t\cup\{s\}\right)} q(e).
\]
Taking the maximum over \(t\) gives the claim.
\end{proof}

For each step \(t=1,\ldots,L\), the site \(v_{t}\) inherits two ordered lists of incoming and outgoing legs
\[
I_{t}=(e_1,\dots,e_r)\quad O_{t}=(f_1,\dots,f_s),
\]
where incoming legs are the incident frontier legs on the processed side, outgoing legs are the incident frontier legs on the unprocessed side, and physical legs keep their prescribed input/output type.

These lists determine the local unfolded map
\[
A_{t}:=A^{(v_{t})},
\qquad
\beta_{t}:=\|A_{t}\|_2,
\qquad
C_{t}:=\beta_{t}^{-1}A_{t},
\]
in the nondegenerate case \(\beta_{t}>0\).

Choose a padded site register
\[
\widetilde{\mathcal H}^{(v_{t})}
\]
of dimension \(\max(m_{t},n_{t})\), together with the canonical isometric embeddings
\[
J_{\mathrm{in}}:\mathcal H_{\mathrm{in}}^{(v_{t})}\hookrightarrow \widetilde{\mathcal H}^{(v_{t})},
\qquad
J_{\mathrm{out}}:\mathcal H_{\mathrm{out}}^{(v_{t})}\hookrightarrow \widetilde{\mathcal H}^{(v_{t})}.
\]
This gives the square embedded contraction
\[
\widetilde C_{t}
=
J_{\mathrm{out}}\,C_{t}\,J_{\mathrm{in}}^\dagger.
\]

The compiler operates on a reusable pool of qubit registers.
At step \(t\), the incident incoming bundle is packed into a local register of size \(\sum_{e\in I_t}q(e)\), padded with zeros to
\[
\max\left(\sum_{e\in I_t}q(e),\sum_{e\in O_t}q(e)\right),
\]
acted on by the synthesized local gadget, and then unpacked as the incident outgoing bundle.
Nonincident frontier registers pass through unchanged.

The packing and unpacking maps here are only canonical tensor-product identifications together with zero-padding of unused qubit slots.
They do not change the selected branch.
Thus, even when
\[
\dim \mathcal H_{\mathrm{in}}^{(v_t)}
\neq
\dim \mathcal H_{\mathrm{out}}^{(v_t)},
\]
the local selected branch realizes the genuine rectangular map
\[
C_t:\mathcal H_{\mathrm{in}}^{(v_t)}\to \mathcal H_{\mathrm{out}}^{(v_t)}
\]
between the incident frontier factors, while padded sectors are initialized to \(\ket{0}\), discarded only after being certified irrelevant to the selected branch, or retained as reusable zero slots.

Now choose any local unitary \(Q_t\) whose selected block on the padded site register is exactly
\[
\widetilde C_t=J_{\mathrm{out}}\,C_t\,J_{\mathrm{in}}^\dagger.
\]
On the selected branch, the packed local input is supported on \(J_{\mathrm{in}}\mathcal H_{\mathrm{in}}^{(v_t)}\), and the local output lies in \(J_{\mathrm{out}}\mathcal H_{\mathrm{out}}^{(v_t)}\).
After unpacking, the selected branch therefore has support exactly on the outgoing frontier factors.
Any local slot outside those factors is certified zero relative to the updated selected set.
By Lemma~\ref{lem:certified-reuse}, its physical slot may therefore be reused immediately as a fresh \(\ket{0}\) ancilla on all later selected branches.

Some local normalized maps are isometries and admit unflagged unitary extensions.
All other local maps use the universal one-flag primitive in the default resource accounting.

For resource bookkeeping, we use the default local realization policy from Section~\ref{sec:local}:
use an unflagged unitary extension when \(C_t\) is an isometry, and otherwise use the universal one-flag primitive.

\begin{definition}[Genuinely dilated steps and \(D(\pi)\)]
Under the default local realization policy of Section~\ref{sec:local}, let \(D(\pi)\) denote the number of steps \(t\) for which the implementation introduces a primitive one-qubit success flag.
Equivalently, \(D(\pi)\) counts the steps for which the normalized local map
\[
C_t:\mathcal H_{\mathrm{in}}^{(v_t)}\to\mathcal H_{\mathrm{out}}^{(v_t)}
\]
is not an isometry.
Thus a strict coisometry has local scale \(1\), but is still counted in \(D(\pi)\) under the default isometry-only shortcut unless it is also an isometry.
\end{definition}

\subsection{Partial contractions and sweep realization}

For each
\[
t=0,1,\dots,L,
\]
let \(H_t\) denote the exact partial contraction obtained after processing the sites in
\[
P_t=\{v_1,\dots,v_t\},
\]
viewed as a map
\[
H_t:\mathcal H(F_0)\to \mathcal H(F_t)
\]
under the canonical frontier orderings.
By construction,
\[
H_0=I,
\qquad
H_L=H(\mathcal T).
\]

Let
\[
\Gamma_t:=\prod_{j=1}^{t}\beta_j,
\qquad
\Gamma_0:=1.
\]

Let
\[
\widehat H_t
\]
denote the selected block of the compiled circuit after the first \(t\) sweep steps.
Thus \(H_t\) and \(\widehat H_t\) have the same source and target frontier spaces.

For step \(t+1\), let
\[
R_t:=F_t\setminus I_{t+1}.
\]
Fix the canonical reorderings
\[
P_t^{\mathrm{in}}:\mathcal H(F_t)\to \mathcal H(I_{t+1})\otimes \mathcal H(R_t),
\qquad
P_t^{\mathrm{out}}:\mathcal H(F_{t+1})\to \mathcal H(O_{t+1})\otimes \mathcal H(R_t),
\]
which separate the incident frontier factors from the untouched remainder.

If \(\beta_j=0\) for some sweep step \(j\), the locally certified zero case is dispatched as in Appendix~\ref{app:zero}.
For the rest of this subsection we assume
\[
\beta_j>0
\qquad
\text{for all }j=1,\dots,L.
\]

\begin{theorem}[Sweep selected-block realization]\label{thm:global-be}
Let \(\mathcal T\) be a TN and let
\[
\pi=(v_1,\dots,v_L)
\]
be any sweep order such that all sweep-induced local scales satisfy \(\beta_t>0\).
At each step \(t\), choose any local unitary \(Q_t\) whose selected block on the padded site register is
\[
\widetilde C_t=J_{\mathrm{out}}\,C_t\,J_{\mathrm{in}}^\dagger,
\]
where \(J_{\mathrm{in}}\) and \(J_{\mathrm{out}}\) are the canonical padding embeddings determined by the sweep.
Then for every
\[
t=0,1,\dots,L,
\]
the partial contraction and the compiled selected block satisfy
\[
H_t=\Gamma_t\,\widehat H_t.
\]
In particular,
\[
H(\mathcal T)=\Gamma(\pi)\,\widehat H_L,
\qquad
\Gamma(\pi)=\Gamma_L=\prod_{t=1}^{L}\beta_t.
\]
\end{theorem}

\begin{proof}
We argue by induction on \(t\).

For \(t=0\), no site has been processed.
Hence both the exact partial contraction and the compiled selected block are the identity on the initial frontier space:
\[
H_0=I=\widehat H_0,
\qquad
\Gamma_0=1.
\]

Now assume
\[
H_t=\Gamma_t\,\widehat H_t
\]
for some \(t<L\).
At step \(t+1\), the sweep determines the unfolded local map and scale
\[
A_{t+1}=A^{(v_{t+1})},
\qquad
\beta_{t+1}=\|A_{t+1}\|_2,
\qquad
C_{t+1}=\beta_{t+1}^{-1}A_{t+1}.
\]
Choose a local unitary \(Q_{t+1}\) whose selected block on the padded site register is
\[
\widetilde C_{t+1}=J_{\mathrm{out}}\,C_{t+1}\,J_{\mathrm{in}}^\dagger.
\]
Since the packed local input is supported on \(J_{\mathrm{in}}\mathcal H_{\mathrm{in}}^{(v_{t+1})}\), the selected branch maps it into \(J_{\mathrm{out}}\mathcal H_{\mathrm{out}}^{(v_{t+1})}\).
After unpacking, the induced action on the genuine frontier factors is exactly \(C_{t+1}\).

The exact partial contraction updates as
\[
H_{t+1}
=
\left(P_t^{\mathrm{out}}\right)^\dagger
\left(A_{t+1}\otimes I_{R_t}\right)
P_t^{\mathrm{in}}\,H_t.
\]
Likewise, the compiled selected block updates as
\[
\widehat H_{t+1}
=
\left(P_t^{\mathrm{out}}\right)^\dagger
\left(C_{t+1}\otimes I_{R_t}\right)
P_t^{\mathrm{in}}\,\widehat H_t.
\]
Therefore
\begin{align*}
\Gamma_{t+1}\widehat H_{t+1}
&=
\left(P_t^{\mathrm{out}}\right)^\dagger
\left(\beta_{t+1}C_{t+1}\otimes I_{R_t}\right)
P_t^{\mathrm{in}}\,\Gamma_t\widehat H_t \\
&=
\left(P_t^{\mathrm{out}}\right)^\dagger
\left(A_{t+1}\otimes I_{R_t}\right)
P_t^{\mathrm{in}}\,H_t \\
&=
H_{t+1}.
\end{align*}
where we used
\[
\beta_{t+1}C_{t+1}=A_{t+1}
\]
and the induction hypothesis, which closes the induction.
\end{proof}

\subsection{Hierarchical online flag aggregation}\label{sec:reuse-log}

A naive implementation would keep all \(D(\pi)\) primitive dilation flags live until the end of the sweep.
This is unnecessary.
Using the binary merge gadget of Lemma~\ref{lem:merge-gadget}, these success conditions can be aggregated online into a logarithmic number of reusable dirty slots.
The level structure is not merely a bookkeeping convenience: a merge target must be certified zero by a flag set disjoint from its controls, since the XOR-type gadget is clean only on a target initialized unconditionally to \(\ket{0}\).
The strict level ordering guarantees this disjointness: every level-\(j\) target is certified by strictly higher levels, which are never level-\(j\) controls.
A naive single running accumulator, by contrast, would reuse a target certified by the accumulator itself, reintroducing off-branch garbage and breaking the selected block.
It is this disjoint-certification structure that yields a logarithmic rather than linear flag count.

We maintain slots arranged in levels, where a level-\(j\) slot represents a block of \(2^{j-1}\) primitive flags.
At each level \(j\), there are up to two standard slots \(a_j,b_j\), and the highest occupied level may additionally carry one extra top slot.
A slot is free if it is certified zero and occupied otherwise, and we maintain the left-packed invariant that occupied slots fill from left to right.

Figure~\ref{fig:flag-merge-tree} illustrates the merge pattern.

\begin{figure}[ht]
\centering
\begin{tikzpicture}[
  x=1mm,y=1mm,
  every node/.style={font=\small,inner sep=1pt},
  dirtyslot/.style={circle, draw=black, fill=black, minimum size=4mm, inner sep=0pt},
  freeslot/.style={circle, draw=black, fill=white, minimum size=4mm, inner sep=0pt},
  vtree/.style={draw=black!45, line width=0.55pt},
  vleaf/.style={circle, draw=black!35, fill=black!10, minimum size=2mm, inner sep=0pt}
]


\def\xa{-8}
\def\xb{8}
\def\yup{7}
\def\ydown{-4}

\def\LX{-30}
\def\RX{30}
\def\TY{24}
\def\BY{-10}


\newcommand{\TwoLeafTreeL}[2]{%
  \draw[vtree] ({#1-2},{#2}) -- ({#1-5},{#2});
  \draw[vtree] ({#1-5},{#2}) -- ({#1-10},{#2+3});
  \draw[vtree] ({#1-5},{#2}) -- ({#1-10},{#2-3});
  \node[vleaf] at ({#1-10},{#2+3}) {};
  \node[vleaf] at ({#1-10},{#2-3}) {};
}

\newcommand{\TwoLeafTreeR}[2]{%
  \draw[vtree] ({#1+2},{#2}) -- ({#1+5},{#2});
  \draw[vtree] ({#1+5},{#2}) -- ({#1+10},{#2+3});
  \draw[vtree] ({#1+5},{#2}) -- ({#1+10},{#2-3});
  \node[vleaf] at ({#1+10},{#2+3}) {};
  \node[vleaf] at ({#1+10},{#2-3}) {};
}

\newcommand{\FourLeafTreeL}[2]{%
  \draw[vtree] ({#1-2},{#2}) -- ({#1-4.5},{#2});
  \draw[vtree] ({#1-4.5},{#2}) -- ({#1-8},{#2+3});
  \draw[vtree] ({#1-4.5},{#2}) -- ({#1-8},{#2-3});

  \draw[vtree] ({#1-8},{#2+3}) -- ({#1-12},{#2+5.4});
  \draw[vtree] ({#1-8},{#2+3}) -- ({#1-12},{#2+1.6});
  \draw[vtree] ({#1-8},{#2-3}) -- ({#1-12},{#2-1.6});
  \draw[vtree] ({#1-8},{#2-3}) -- ({#1-12},{#2-5.4});

  \node[vleaf] at ({#1-12},{#2+5.4}) {};
  \node[vleaf] at ({#1-12},{#2+1.6}) {};
  \node[vleaf] at ({#1-12},{#2-1.6}) {};
  \node[vleaf] at ({#1-12},{#2-5.4}) {};
}

\newcommand{\FourLeafTreeR}[2]{%
  \draw[vtree] ({#1+2},{#2}) -- ({#1+4.5},{#2});
  \draw[vtree] ({#1+4.5},{#2}) -- ({#1+8},{#2+3});
  \draw[vtree] ({#1+4.5},{#2}) -- ({#1+8},{#2-3});

  \draw[vtree] ({#1+8},{#2+3}) -- ({#1+12},{#2+5.4});
  \draw[vtree] ({#1+8},{#2+3}) -- ({#1+12},{#2+1.6});
  \draw[vtree] ({#1+8},{#2-3}) -- ({#1+12},{#2-1.6});
  \draw[vtree] ({#1+8},{#2-3}) -- ({#1+12},{#2-5.4});

  \node[vleaf] at ({#1+12},{#2+5.4}) {};
  \node[vleaf] at ({#1+12},{#2+1.6}) {};
  \node[vleaf] at ({#1+12},{#2-1.6}) {};
  \node[vleaf] at ({#1+12},{#2-5.4}) {};
}

\newcommand{\SlotLabels}{%
  \node[font=\scriptsize] at (\xa,12) {\(a_{j+1}\)};
  \node[font=\scriptsize] at (\xb,12) {\(b_{j+1}\)};
  \node[font=\scriptsize] at (\xa,1) {\(a_j\)};
  \node[font=\scriptsize] at (\xb,1) {\(b_j\)};
}


\begin{scope}[shift={(\LX,\TY)}]
  \node at (0,16) {\textbf{(a)}};
  \SlotLabels

  \node[freeslot] at (\xa,\yup) {};
  \node[freeslot] at (\xb,\yup) {};

  \node[dirtyslot] at (\xa,\ydown) {};
  \node[dirtyslot] at (\xb,\ydown) {};

  \TwoLeafTreeL{\xa}{\ydown}
  \TwoLeafTreeR{\xb}{\ydown}
\end{scope}


\begin{scope}[shift={(\RX,\TY)}]
  \node at (0,16) {\textbf{(b)}};
  \SlotLabels

  \node[dirtyslot] at (\xa,\yup) {};
  \node[freeslot] at (\xb,\yup) {};

  \node[freeslot] at (\xa,\ydown) {};
  \node[freeslot] at (\xb,\ydown) {};

  \FourLeafTreeL{\xa}{\yup}
\end{scope}


\begin{scope}[shift={(\LX,\BY)}]
  \node at (0,16) {\textbf{(c)}};
  \SlotLabels

  \node[dirtyslot] at (\xa,\yup) {};
  \node[freeslot] at (\xb,\yup) {};

  \node[dirtyslot] at (\xa,\ydown) {};
  \node[dirtyslot] at (\xb,\ydown) {};

  \FourLeafTreeL{\xa}{\yup}
  \TwoLeafTreeL{\xa}{\ydown}
  \TwoLeafTreeR{\xb}{\ydown}
\end{scope}


\begin{scope}[shift={(\RX,\BY)}]
  \node at (0,16) {\textbf{(d)}};
  \SlotLabels

  \node[dirtyslot] at (\xa,\yup) {};
  \node[dirtyslot] at (\xb,\yup) {};

  \node[freeslot] at (\xa,\ydown) {};
  \node[freeslot] at (\xb,\ydown) {};

  \FourLeafTreeL{\xa}{\yup}
  \FourLeafTreeR{\xb}{\yup}
\end{scope}


\node at (0,\TY+1) {\(\mapsto\)};
\node at (0,\BY+1) {\(\mapsto\)};

\end{tikzpicture}
\caption{Snapshots of the online flag-merge hierarchy.
Black circles denote occupied dirty slots and white circles denote free slots.
Each gray tree records the primitive flags represented by that slot.}
\label{fig:flag-merge-tree}
\end{figure}

Operationally, before a local dilation writes a new primitive flag into level \(1\), the hierarchy first performs any required upward merges until a certified-zero level-\(1\) slot is available.
The new primitive flag is then written into that slot.
The slot pool is exactly the set of flag qubits currently used to hold primitive or merged success conditions; no separate permanently live primitive-flag register is kept.

\begin{lemma}[Exactness of hierarchical aggregation]\label{lem:reuse-hierarchy}
Consider any online merge schedule built from repeated applications of Lemma~\ref{lem:merge-gadget}, under the left-packed invariant.
After any number \(t\) of inserted primitive flags:
\begin{enumerate}
\item each occupied level-\(j\) slot represents exactly \(2^{j-1}\) primitive flags;
\item every free slot below the highest occupied level is certified zero by occupied slots at higher levels; and
\item projecting all occupied slots to \(0\) is exactly equivalent to projecting all \(t\) primitive flags to \(0\).
\end{enumerate}
\end{lemma}

\begin{proof}
Induct on \(t\).

The claim is trivial for \(t=0\).
Assume it holds after \(t\) insertions and insert one more primitive flag.

If the new flag occupies a free level-\(1\) slot, nothing else changes.
Otherwise, insertion triggers repeated upward merges whenever the current level has no free slot.
Each merge replaces two occupied level-\(j\) slots, together representing
\[
2^{j-1}+2^{j-1}=2^j
\]
primitive flags, by one occupied level-\((j+1)\) slot representing exactly that same block.
The selected branch is preserved exactly, and the consumed child slots become certified zero relative to the new parent slot.
Thus all three claims remain true.
In particular, later reuse of a free slot is exact by Lemma~\ref{lem:certified-reuse}.
\end{proof}

\begin{definition}[Slot capacity]
For \(s\ge 0\), let \(\mathrm{cap}(s)\) denote the maximum number of primitive flags that can be represented exactly by this left-packed hierarchy using \(s\) slots.
\end{definition}

\begin{proposition}[Capacity formula]\label{prop:slot-capacity}
For \(s\ge 0\),
\[
\mathrm{cap}(s)=
\begin{cases}
2^{s/2+1}-2, & \text{if } s \text{is even},\\[1mm]
3\cdot 2^{(s-1)/2}-2, & \text{if }s\text{ is odd}.
\end{cases}
\]
\end{proposition}

\begin{proof}
By Lemma~\ref{lem:reuse-hierarchy}, a level-\(j\) slot represents \(2^{j-1}\) primitive flags.

With \(2k\) slots, the maximal left-packed arrangement places two slots on each level \(1,\dots,k\).
Its total represented capacity is therefore
\[
2\sum_{j=1}^{k}2^{j-1}=2^{k+1}-2.
\]

With \(2k+1\) slots, one may additionally place one more slot at level \(k+1\), which contributes \(2^k\) further primitive flags.
Hence
\[
\mathrm{cap}(2k+1)=\left(2^{k+1}-2\right)+2^k=3\cdot 2^k-2.
\]
\end{proof}

\begin{corollary}[Online flag aggregation bound]\label{cor:reuse-log}
The online bookkeeping can be performed using
\[
\min\{s\ge 0:\; D(\pi)\le \mathrm{cap}(s)\}
\]
additional flag qubits.
In particular, this requires only \(\bigO{\log(D(\pi)+1)}\) additional flag qubits.

Moreover, the number of online merge gadgets is at most
\[
D(\pi)-1
\]
when \(D(\pi)\ge 1\), and \(0\) when \(D(\pi)=0\).
\end{corollary}

\begin{proof}
By Proposition~\ref{prop:slot-capacity}, a left-packed hierarchy with \(s\) slots can represent exactly \(\mathrm{cap}(s)\) primitive flags.
Hence the online bookkeeping can be performed using
\[
\min\{s\ge 0:\; D(\pi)\le \mathrm{cap}(s)\}
\]
additional flag qubits.

For \(D(\pi)\ge 1\), Proposition~\ref{prop:slot-capacity} implies that this quantity is \(\bigO{\log D(\pi)}\).
If \(D(\pi)=0\), no additional flag qubit is needed.

For the merge count, if \(D(\pi)=0\) there are no merges.
If \(D(\pi)\ge 1\), each primitive insertion increases the number of represented blocks by \(1\), while each merge decreases it by \(1\).
After \(D(\pi)\) insertions, at least one represented block remains.
Hence the total number of merges is at most \(D(\pi)-1\).
\end{proof}

\section{Compiler resources, sweep dependence, and bounded-local correspondence}\label{sec:complexity}

Fix a TN \(\mathcal T\) and a sweep \(\pi=(v_1,\dots,v_L)\).
Unless stated otherwise, this section concerns the nondegenerate sweep branch \(\beta_t>0\) for all \(t\); the locally certified zero branch returns the canonical zero BE as in Appendix~\ref{app:zero}.
Section~\ref{sec:global} associates to this sweep an explicit BE of \(H(\mathcal T)\) with scale \(\Gamma(\pi)\), frontier memory \(M(\pi)\), and \(D(\pi)\) genuinely dilated local steps.
We now quantify compile time, memory, sweep-dependent loss, and the bounded-local specialization.

\subsection{Fixed-sweep compiler resources}

\begin{theorem}[Fixed-sweep compile complexity]\label{thm:compile}
For each site \(v\in V\), let
\[
A^{(v)}\in\C^{m_v\times n_v}
\]
be the unfolded site matrix induced by \(\pi\), and set
\[
k_v:=\max(m_v,n_v).
\]
Then the compiler of Sections~\ref{sec:local} and~\ref{sec:global} constructs the qubit BE specified by Theorem~\ref{thm:global-be} using a number of arithmetic operations polynomial in the explicit input length of \(\mathcal T\).

More concretely, one dense route has arithmetic cost
\[
T_{\mathrm{compile}}(\mathcal T,\pi)
\in
\bigO{
\sum_{v\in V}\left(\min(m_v,n_v)\,m_v n_v + k_v^3\right)
},
\]
up to lower-order sweep bookkeeping and register-routing metadata.
The synthesized local circuit at site \(v\) acts on
\[
q_v=\left\lceil \log_2 k_v\right\rceil+\bigO{1}
\]
qubits and has \(\bigO{k_v^2}\) one- and two-qubit gates under standard dense synthesis.
\end{theorem}

\begin{proof}
For each site \(v\), the compiler unfolds \(T^{(v)}\) into \(A^{(v)}\), computes
\[
\beta_v=\|A^{(v)}\|_2,
\]
forms the normalized local contraction, chooses a local selected-block realization, and synthesizes the resulting local unitary on the padded site register.
A dense singular-value decomposition contributes
\[
\bigO{\min(m_v,n_v)\,m_v n_v}
\]
arithmetic operations.
The remaining local dense linear algebra and local synthesis depend only on the padded local dimension \(k_v\).
For one standard dense route, this contributes the \(k_v^3\) term, while the synthesized local gate count is \(\bigO{k_v^2}\).

Since
\[
k_v\le m_v n_v=|T^{(v)}|,
\]
every local term is polynomial in the explicit local tensor size.
Summing over \(v\in V\), and adding the sweep and incidence bookkeeping, proves that the total compile time is polynomial in the explicit input length.
The stated local gate count is the standard dense synthesis bound for a \(q_v\)-qubit unitary.
\end{proof}

\begin{corollary}[Compiled resource profile]\label{cor:compiled-profile}
The compiler produces a BE of \(H(\mathcal T)\) with:
\begin{itemize}
\item scale \(\Gamma(\pi)\),
\item sweep memory \(M(\pi)\),
\item an online dirty-flag slot pool of size \(\bigO{\log(D(\pi)+1)}\), and hence also \(\bigO{\log(L+1)}\),
\item at most \(D(\pi)-1\) online flag-merge gadgets if \(D(\pi)\ge 1\), and none if \(D(\pi)=0\).
\end{itemize}
Thus the total memory is
\[
M(\pi)+\bigO{\log(D(\pi)+1)},
\]
with the logarithmic term accounting exactly for the reusable dirty-flag slot pool.
\end{corollary}

\begin{proof}
By Theorem~\ref{thm:global-be}, the compiler produces a BE with scale \(\Gamma(\pi)\) representing \(H(\mathcal T)\).
The sweep memory cost is \(M(\pi)\) by definition of the sweep frontier.
The bookkeeping bound and merge count follow from Corollary~\ref{cor:reuse-log}.
Finally,
\[
D(\pi)\le L.
\]
\end{proof}

\begin{remark}
Here and below, ``memory'' means frontier qubits plus dirty flag slots.
It excludes transient local synthesis or routing workspace.
\end{remark}

\subsection{Sweep dependence and approximation transfer}

For a fixed explicit TN representation \(\mathcal T\), different sweeps can change \(\Gamma(\pi)\), \(M(\pi)\), and \(D(\pi)\).
Exact or approximate preprocessing may first replace \(\mathcal T\) by a different explicit representation of the same map and then change these quantities again.

Before varying the sweep, it is useful to record a sweep-independent floor on \(M(\pi)\).
By Proposition~\ref{prop:cutwidth}, for a fixed augmented graph, \(M(\pi)\) is the weight of the edges crossing the sweep cut, so
\[
\min_{\pi} M(\pi)
\]
is exactly the weighted cutwidth of the augmented network graph under the sweep order.
Equivalently, splitting each multi-qubit bond into that many parallel single-qubit edges, it is the (unweighted) cutwidth of the resulting multigraph.
The frontier decomposes into a bond-structural part and a global-leg part.
Let \(M_{\mathrm{bond}}(\pi)\) count the qubits on bond legs crossing the sweep cut, and let
\[
G_{\mathrm{in}}:=\sum_{v}\sum_{\text{input legs }\ell\text{ at }v}q(\ell),
\qquad
G_{\mathrm{out}}:=\sum_{v}\sum_{\text{output legs }\ell\text{ at }v}q(\ell)
\]
be the total qubit weights of all global input and output legs.
The global input legs on unprocessed sites and global output legs on processed sites are monotone in \(t\) and contribute at least \(\max\{G_{\mathrm{in}},G_{\mathrm{out}}\}\), a layout-independent additive floor.
Since the vertex separation number of a graph equals its pathwidth \cite{Kinnersley1992}, and each frontier vertex of degree at most \(\Delta\) contributes between one and \(\Delta\) crossing edges (each of weight at most \(q_{\max}\)), the bond-structural part satisfies the sandwich
\[
\operatorname{pw}(G)\;\le\;\min_{\pi}M_{\mathrm{bond}}(\pi)\;\le\;\Delta\,q_{\max}\,\operatorname{pw}(G).
\]
In the bounded-local regime \(\Delta=k_{\max}\) and \(q_{\max}\) are constants, so
\[
\min_{\pi}M(\pi)=\Theta\!\left(\operatorname{pw}(G)+G_{\mathrm{in}}+G_{\mathrm{out}}\right):
\]
weighted pathwidth is the sweep-independent floor on the structural frontier memory, tight up to the same constant factor as the bounded-local model itself, with the global-leg totals entering only as an additive layout-independent floor.
(For subcubic unweighted graphs the two parameters coincide up to an additive constant; in fact \(\operatorname{cw}=\operatorname{pw}+1\).)

This separates the two levers the interface exposes:
classical TN preprocessing that changes the network geometry, such as rerouting, flattening, or series-parallel and triangle reduction, can lower the achievable frontier memory by lowering the weighted pathwidth before compilation, while the choice of sweep only realizes a cutwidth budget that cannot drop below it.
Treewidth is a weaker lower bound on the same quantity and is loose by a logarithmic factor even in bounded-local graphs; we therefore use pathwidth as the sweep-independent floor.

Set
\[
\Gamma_{\mathrm{glob}}:=\|H(\mathcal T)\|_2,
\qquad
\rho(\pi):=\frac{\Gamma_{\mathrm{glob}}}{\Gamma(\pi)}.
\]
Let \(\widehat H_\pi\) denote the selected block of the compiled BE, so that
\[
H(\mathcal T)=\Gamma(\pi)\,\widehat H_\pi.
\]
Since \(\widehat H_\pi\) is a selected block of a unitary, it is a contraction.
Hence
\[
\|H(\mathcal T)\|_2\le \Gamma(\pi),
\]
and therefore
\[
0\le \rho(\pi)\le 1.
\]

\begin{proposition}[Post-selection factorization]\label{prop:post-factor}
For any normalized input state \(x\),
\[
p_\pi(x)=\|\widehat H_\pi x\|^2
=\frac{\|H(\mathcal T)x\|^2}{\Gamma(\pi)^2}.
\]
If \(\Gamma_{\mathrm{glob}}>0\), then
\[
p_\pi(x)
=
\left\|\Gamma_{\mathrm{glob}}^{-1}H(\mathcal T)x\right\|^2 \rho(\pi)^2.
\]
\end{proposition}

\begin{proof}
The first equality is by definition of selected-branch success probability,
\[
p_\pi(x)
=
\|\widehat H_\pi x\|^2.
\]

If \(\Gamma_{\mathrm{glob}}>0\), then
\[
\|\widehat H_\pi x\|^2
=
\|\Gamma(\pi)^{-1}H(\mathcal T)x\|^2
=
\|\Gamma_{\mathrm{glob}}^{-1}H(\mathcal T)x\|^2\rho(\pi)^2.
\]
\end{proof}

The meaningful sweep-dependent loss is therefore \(\rho(\pi)\), rather than an input-independent success probability.
Section~\ref{sec:structured} identifies the optimal case \(\rho(\pi)=1\).

A small worked example illustrating sweep-dependent scale and frontier memory, and the round-trip restoration of a scale loss introduced by a bad sweep, is given in Appendix~\ref{app:toy-example}.

Classical TN preprocessing can also be performed before recompilation.
The main point is that local tensor replacements induce controlled operator error.

\begin{proposition}[Lipschitz stability under local replacement]\label{prop:lipschitz}
Consider two TNs on the same graph and with the same leg structure,
\[
\mathcal T=\{T^{(v)}\}_{v\in V},
\qquad
\widetilde{\mathcal T}=\{\widetilde T^{(v)}\}_{v\in V},
\]
where the unfolded site operators are taken with respect to the same sweep \(\pi\).
Let \(A^{(v)}\) and \(\widetilde A^{(v)}\) denote the corresponding unfolded site operators with respect to the same sweep \(\pi\).
Write
\[
A_t:=A^{(v_t)},
\qquad
\widetilde A_t:=\widetilde A^{(v_t)},
\qquad
\beta_t:=\|A_t\|_2,
\qquad
\widetilde\beta_t:=\|\widetilde A_t\|_2.
\]
Then
\[
\bigl\|H(\mathcal T)-H(\widetilde{\mathcal T})\bigr\|_2
\le
\sum_{t=1}^{L}
\left(
\prod_{j<t}\beta_j
\right)
\|A_t-\widetilde A_t\|_2
\left(
\prod_{j>t}\widetilde\beta_j
\right).
\]
In particular, if
\[
\beta_t,\widetilde\beta_t\le M
\qquad\text{for all }t,
\]
then
\[
\bigl\|H(\mathcal T)-H(\widetilde{\mathcal T})\bigr\|_2
\le
M^{L-1}\sum_{t=1}^{L}\|A_t-\widetilde A_t\|_2.
\]
\end{proposition}

\begin{proof}
See Appendix~\ref{app:lipschitz-proof}.
\end{proof}

A useful special case is replacement of a near-isometric local tensor by its polar factor.

\begin{proposition}[Polar replacement]\label{prop:polar}
Let
\[
A\in\C^{m\times n}
\]
and let
\[
A=WP
\]
be its polar decomposition.
Then \(W\) is exactly isometric on \((\ker A)^\perp\), and after restricting the domain to \((\ker A)^\perp\),
\[
\|A-W\|_2=\|P-I\|_2.
\]
\end{proposition}

\begin{proof}
On the support of \(A\), the partial isometry \(W\) is an isometry and
\[
A-W=W(P-I).
\]
Therefore
\[
\|A-W\|_2=\|P-I\|_2.
\]
\end{proof}

Together, Propositions~\ref{prop:lipschitz} and~\ref{prop:polar} show that exact or approximate TN preprocessing can be transferred to BE recompilation with explicit operator-norm control.

\subsection{Bounded-local specialization and consequences}

\begin{theorem}[Bounded-local linear compilation]\label{thm:bounded-linear}
Assume every site has degree at most \(k_{\max}\), and every leg dimension is bounded by a constant \(D_{\max}\).
Then:
\begin{itemize}
\item the compile time is
\[
T_{\mathrm{compile}}(\mathcal T,\pi)\in \bigO{L},
\qquad
\text{hence also }
T_{\mathrm{compile}}(\mathcal T,\pi)\in \bigO{\mathrm{size}(\mathcal T)},
\]
\item each compiled local gadget acts on \(\bigO{1}\) qubits, and
\item the total BE circuit size is
\[
\bigO{L},
\]
over arbitrary one- and two-qubit gates.
\end{itemize}
\end{theorem}

\begin{proof}
Bounded locality implies that every site tensor has constant size, so each unfolded matrix has
\[
m_v,n_v\le D_{\max}^{k_{\max}}\in\bigO{1}.
\]
Hence each local preprocessing and local synthesis step takes constant time and the total compile time is \(\bigO{L}\).
Since
\[
\mathrm{size}(\mathcal T)=\Theta(L),
\]
this is also \(\bigO{\mathrm{size}(\mathcal T)}\).

After qubit padding, each leg occupies \(\bigO{1}\) qubits and each site has \(\bigO{1}\) incident legs, so each compiled local gadget acts on \(\bigO{1}\) qubits.
There is one compiled local gadget per site and at most \(D(\pi)-1\in\bigO{L}\) merge gadgets by Corollary~\ref{cor:reuse-log}.
Each merge gadget acts on three qubits and decomposes into \(\bigO{1}\) one- and two-qubit gates, so merge gadgets also contribute only \(\bigO{L}\) gates in total.

Each compiled local gadget acts on \(\bigO{1}\) qubits, so after standard local synthesis it contributes \(\bigO{1}\) one- and two-qubit gates.
Since there are \(L\) compiled local gadgets and at most \(D(\pi)-1\in\bigO{L}\) merge gadgets, the total BE circuit size is \(\bigO{L}\).
\end{proof}

\begin{remark}
If a fixed finite universal gate set is imposed afterwards, the usual exact or approximate local synthesis overhead is additional.
In the bounded-local regime this overhead is per constant-size local gadget.
\end{remark}

\begin{theorem}[Bounded-local TN/BE correspondence for arbitrary finite linear maps]\label{thm:correspondence}
In the unit-cost arithmetic model of Section~\ref{sec:results}, bounded-local TN representations and bounded-local BE representations correspond up to constant-factor overhead in the following sense.

For every fixed pair of TN locality constants \(q_{\max}\) and \(k_{\max}\), there is a constant \(c_1\) such that any TN of size \(T\) whose sites have degree at most \(k_{\max}\) and whose padded legs use at most \(q_{\max}\) qubits compiles to a bounded-local BE with
\[
\mathrm{size}_{\mathrm{BE}}\le c_1 T.
\]

Conversely, there is a constant \(c_2\) such that every bounded-local BE of size \(T\) canonically yields a bounded-local TN of size at most \(c_2 T\).
\end{theorem}

\begin{proof}
The TN-to-BE direction is exactly Theorem~\ref{thm:bounded-linear}.

For the BE-to-TN direction, replace each one-qubit gate by its degree-two tensor and each two-qubit gate by its degree-four tensor.
Circuit wires become internal bonds of dimension \(2\), with degree-two identity tensors inserted for pass-through wires when needed.
Input and output register declarations become free input and output legs.
Initialized ancillas and post-selected ancillas are imposed as fixed boundary values on the corresponding circuit indices; equivalently, one may contract with constant-size boundary ket and bra tensors.
Thus every gate, wire, and boundary condition contributes only constant-size data, and the total number of tensors and bonds is \(\bigO{T}\).

The resulting TN contracts to the selected block \(B=\alpha^{-1}H\).
Absorb the scalar \(\alpha\) into any existing tensor; if there is no tensor, add one degree-zero scalar tensor with entry \(\alpha\).
This changes only a constant-size tensor entry, or adds one constant-size tensor, and does not affect the asymptotic size or locality.
\end{proof}

\begin{corollary}[Selected-block round trip]\label{cor:round-trip}
Let a bounded-local BE of size \(T\) represent
\[
H=\alpha B.
\]
Turn its gates, initialized ancillas, post-selected ancillas, and input/output register pattern into the standard circuit TN for the selected block \(B\), imposing preparation and post-selection as fixed boundary values.
Absorb the scalar \(\alpha\) into one existing tensor, adding one degree-zero scalar tensor if no tensor exists.

With the boundary-aware convention that BE preparation and post-selection data remain boundary data under recompilation, the inherited circuit-time order recompiles this TN to a BE for \(H\) with scale \(\alpha\) and size \(\bigO{T}\), provided no compression or refactorization is performed.
\end{corollary}

\begin{proof}
Every circuit gate tensor unfolds as a unitary in the inherited circuit-time order.
Hence gate tensors contribute local scale \(1\).
Preparation and post-selection are fixed boundary values defining the selected block, so they do not create additional local scale factors.
The scalar-absorbing tensor, or the added scalar tensor if necessary, contributes scale \(\alpha\).

Therefore the sweep composition theorem gives selected block \(B\) and represented operator
\[
H=\alpha B.
\]
All tensors and boundary data have constant size, and the inherited circuit TN has \(\bigO{T}\) total size, so recompilation has size \(\bigO{T}\).
\end{proof}

\begin{corollary}[Monotone round trip under faithful Schmidt-rank compression]\label{cor:round-trip-compress}
Let a bounded-local BE of size \(T\) represent
\[
H=\alpha B.
\]
Turn its gates, initialized ancillas, post-selected ancillas, and input/output register pattern into the standard circuit TN for the selected block \(B\), imposing preparation and post-selection as fixed boundary values.
Apply faithful Schmidt-rank compression to the resulting circuit TN, i.e.\ replace each bond by its dominant Schmidt subspace, and re-canonicalize the local tensors accordingly.
Recompile the compressed TN in the inherited circuit-time order.

Then the recompiled BE represents \(H'=\alpha B'\) with
\[
\Gamma'\le 1,
\qquad
\alpha' = \alpha\,\Gamma' \le \alpha,
\qquad
M'\le M,
\qquad
\|H-H'\|_2 \le \alpha\sum_{v\in V}\bigl\|A^{(v)}-\widetilde A^{(v)}\bigr\|_2,
\]
where \(\Gamma=1\) is the scale, \(M\) the frontier memory, and \(\alpha\) the scale factor of the identity recompilation (Corollary~\ref{cor:round-trip}) before compression (so \(\Gamma'\le 1\) and \(\alpha'=\alpha\,\Gamma'\) are the post-compression values), each \(A^{(v)}\) is the inherited unitary unfolding (\(\|A^{(v)}\|_2=1\)), each \(\widetilde A^{(v)}\) its truncated counterpart (\(\|\widetilde A^{(v)}\|_2\le 1\)), and the last sum is bounded by the total discarded Schmidt weight across all bonds.

In particular, faithful compression never increases the scale cost or the frontier memory, and transfers to the recompiled BE with explicit operator-norm control.
\end{corollary}

\begin{proof}
In the inherited circuit-time order, every gate tensor unfolds as a unitary, so \(\beta_t=1\) and the identity round trip of Corollary~\ref{cor:round-trip} gives \(\Gamma=1\).
Schmidt-rank compression replaces each bond by an orthogonal projection onto its dominant subspace.
Each truncated local tensor is the original unitary unfolding composed with such projections on its incident bonds, hence is a contraction:
\[
\|\widetilde A^{(v)}\|_2\le \|A^{(v)}\|_2=1.
\]
Therefore every recompiled local scale satisfies \(\widetilde\beta_t\le 1\), and Theorem~\ref{thm:global-be} gives
\[
\Gamma'=\prod_{t=1}^{L}\widetilde\beta_t\le 1,
\qquad
\alpha'=\alpha\,\Gamma'\le \alpha.
\]
Bond dimensions only shrink under truncation, so the frontier memory is monotone non-increasing: \(M'\le M\).
For the error bound, apply Proposition~\ref{prop:lipschitz} to the pair \((\mathcal T,\widetilde{\mathcal T})\) in the inherited order.
Since both \(\beta_t=1\) and \(\widetilde\beta_t\le 1\), the bounded-scale prefactor of that proposition equals \(1\), yielding
\[
\bigl\|H(\mathcal T)-H(\widetilde{\mathcal T})\bigr\|_2
\le
\sum_{t=1}^{L}\bigl\|A_t-\widetilde A_t\bigr\|_2.
\]
Each local difference is the discarded Schmidt weight on the incident bonds of site \(v_t\), and the selected block of the original BE carries scale \(\alpha\), giving the stated bound.
\end{proof}

\begin{remark}[Three regimes of the round trip]\label{rem:round-trip-regimes}
The round trip separates into three regimes with sharply different guarantees.
\begin{enumerate}
\item \emph{Identity.}
With no TN-side modification, recompilation is exact with scale \(\alpha\) and size \(\bigO{T}\) (Corollary~\ref{cor:round-trip}).
\item \emph{Faithful compression.}
Schmidt-rank compression is monotone: \(\alpha'\le\alpha\), \(M'\le M\), with operator error bounded by the discarded weight (Corollary~\ref{cor:round-trip-compress}).
\item \emph{Arbitrary restructuring.}
Gauge changes, reordering, or non-truncation edits admit no general scale guarantee; Proposition~\ref{prop:main-hardness} shows this is unavoidable, since certifying exact scale optimality is already intractable for diagonal MPOs on a path.
\end{enumerate}
Regimes~1 and 2 are the provable core of selected-block optimization; regime~3 is where practical heuristics live, without a general theorem.
\end{remark}

\begin{corollary}[No-go within the explicit bounded-local model]
If an operator family does not admit polynomial-size bounded-local TNs, then it does not admit polynomial-size explicit bounded-local BEs in this model.
\end{corollary}

\begin{proof}
Immediate from Theorem~\ref{thm:correspondence}.
\end{proof}

The selected-block optimization interpretation of this correspondence is discussed in Section~\ref{sec:discussion}.

\section{Scale-optimal sweeps and barriers}\label{sec:structured}

We now isolate the exact condition for scale-optimality and identify a structured exact preprocessing regime that guarantees scale-optimal sweeps.

The key positive statement is an exact preprocessing result for one-sided rooted trees.
After exact Schmidt-rank compression and standard rooted canonicalization, such a tree admits a compatible partial sweep whose compiled partial contraction is scale-optimal.
Under exact Schmidt-rank compression, every bond is reduced to its dominant Schmidt support and the local tensors become isometric or coisometric gauges that carry no local scale loss; the only residual non-isometric content is the part of the network not collapsible into such a gauge.
Bridge-hourglass forests are precisely the structural case in which this residual concentrates at a single bridge site per connected component: all non-bridge sites become norm-preserving (isometric or coisometric) after compression, so the entire compiled scale reduces to the operator norm at the bridge, which equals the global operator norm.
For this class, scale-optimality is therefore not merely achievable but constructive, and it identifies the natural boundary of what exact preprocessing guarantees: whenever the non-isometric residual after compression spans more than a single site per component, no general scale-optimality guarantee is available, and Proposition~\ref{prop:main-hardness} shows this is unavoidable.

A complete characterization of which TNs admit scale-optimal sweeps is not given here; the hourglass family is a broad and natural sufficient class, not a maximality claim.

\subsection{Scale-optimal sweeps}

We first record the terminology and then prove Proposition~\ref{prop:main-scale}.

\begin{definition}[Scale-optimal sweep]
A sweep \(\pi\) is called \emph{scale-optimal} if either \(H(\mathcal T)=0\), or its compiled selected block \(\widehat H_\pi\) satisfies
\[
\|\widehat H_\pi\|_2=1.
\]
For \(H(\mathcal T)\neq 0\), this is equivalent to
\[
\left\|\Gamma(\pi)^{-1}H(\mathcal T)\right\|_2=1.
\]
\end{definition}

\begin{proof}[Proof of Proposition~\ref{prop:main-scale}]
By definition,
\[
\widehat H_\pi=\Gamma(\pi)^{-1}H(\mathcal T),
\]
so
\[
\|H(\mathcal T)\|_2=\Gamma(\pi)\,\|\widehat H_\pi\|_2.
\]
Therefore
\[
\Gamma(\pi)=\|H(\mathcal T)\|_2
\iff
\|\widehat H_\pi\|_2=1.
\]
\end{proof}

\subsection{Optimal preprocessing of one-sided rooted trees}\label{sec:force-canonical}

We use the following standard rooted tree canonicalization fact; see, for example, \cite{Schollwoeck2011,Orus2014,Shi2006,Murg2010}.

\begin{lemma}[Rooted tree canonicalization]\label{lem:rooted-canonical}
Let \(R\) be an explicit tree TN with root \(r\).
By sweeping from the leaves toward \(r\), performing exact SVD-based Schmidt-rank compression on every parent-facing bond, and fixing the standard rooted canonical gauge by absorbing the nonisometric factors toward the root, one obtains in polynomial arithmetic time an equivalent rooted TN in which each compressed parent-facing bond is the exact Schmidt support of the corresponding descendant subtree.
With the orientation toward the root, the non-root unfoldings are coisometries.
With the dual orientation away from the root, they are isometries.
Zero Schmidt sectors are removed, and if the remaining root support is zero then the represented tree contraction is zero.
\end{lemma}

For the compiler, the support alignment in Lemma~\ref{lem:rooted-canonical} is what makes the tree-side contractions scale-optimal.

Figure~\ref{fig:tree-root-sweep} illustrates the input-side case.
The output-side case is the exact dual.

\begin{figure}[ht]
\centering
\begin{tikzpicture}[x=1mm,y=1mm,every node/.style={font=\small,inner sep=1pt}]

\def\LX{-30}
\def\RX{30}

\newcommand{\TreeCoords}{%
  \coordinate (r) at (0,12);
  \coordinate (a) at (-12,0);
  \coordinate (b) at (12,0);
  \coordinate (c) at (-18,-13);
  \coordinate (d) at (-6,-14);
  \coordinate (e) at (18,-13);
}

\newcommand{\TreeNodes}{%
  \node[generaltensor,double,double distance=0.5pt,minimum width=7mm] (Nr) at (r) {\(r\)};
  \node[generaltensor, minimum width=7mm] (Na) at (a) {};
  \node[generaltensor, minimum width=7mm] (Nb) at (b) {};
  \node[generaltensor, minimum width=7mm] (Nc) at (c) {};
  \node[generaltensor, minimum width=7mm] (Nd) at (d) {};
  \node[generaltensor, minimum width=7mm] (Ne) at (e) {};
}

\newcommand{\BondTreeUndirected}{%
  \draw[sw_und] (Nc) -- (Na);
  \draw[sw_und] (Nd) -- (Na);
  \draw[sw_und] (Na) -- (Nr);
  \draw[sw_und] (Ne) -- (Nb);
  \draw[sw_und] (Nb) -- (Nr);
}

\newcommand{\BondTreeDirectedInward}{%
  \draw[sw_active] (Nc) -- (Na);
  \draw[sw_active] (Nd) -- (Na);
  \draw[sw_active] (Na) -- (Nr);
  \draw[sw_active] (Ne) -- (Nb);
  \draw[sw_active] (Nb) -- (Nr);
}

\newcommand{\PhysicalInputsUndirected}{%
  \draw[sw_und] ($(Nc)+(-8,-4)$) -- (Nc);
  \node[font=\scriptsize, anchor=east] at ($(Nc)+(-9,-4)$) {in};

  \draw[sw_und] ($(Nd)+(0,-9)$) -- (Nd);
  \node[font=\scriptsize, anchor=north] at ($(Nd)+(0,-10)$) {in};

  \draw[sw_und] ($(Nb)+(9,6)$) -- (Nb);
  \node[font=\scriptsize, anchor=west] at ($(Nb)+(10,6)$) {in};

  \draw[sw_und] ($(Ne)+(9,-2)$) -- (Ne);
  \node[font=\scriptsize, anchor=west] at ($(Ne)+(10,-2)$) {in};
}

\newcommand{\PhysicalInputsDirected}{%
  \draw[dimleg] ($(Nc)+(-8,-4)$) -- (Nc);
  \draw[dimleg] ($(Nd)+(0,-9)$) -- (Nd);
  \draw[dimleg] ($(Nb)+(9,6)$) -- (Nb);
  \draw[dimleg] ($(Ne)+(9,-2)$) -- (Ne);
}

\newcommand{\RootExteriorUndirected}{%
  \draw[sw_und] (Nr) -- ++(-10,10);
  \draw[sw_und] (Nr) -- ++(10,10);
}

\newcommand{\RootExteriorDirected}{%
  \draw[sw_und] (Nr) -- ++(-10,10);
  \draw[sw_und] (Nr) -- ++(10,10);
}

\begin{scope}[shift={(\LX,0)}]
  \TreeCoords
  \node[anchor=west] at (-25,27) {\textbf{(a) input-one-sided rooted region}};
  \TreeNodes
  \BondTreeUndirected
  \PhysicalInputsUndirected
  \RootExteriorUndirected
\end{scope}

\begin{scope}[shift={(\RX,0)}]
  \TreeCoords
  \node[anchor=west] at (-23,27) {\textbf{(b) compatible inward sweep}};
  \TreeNodes
  \BondTreeDirectedInward
  \PhysicalInputsDirected
  \RootExteriorDirected
\end{scope}

\end{tikzpicture}
\caption{Input-one-sided rooted region and its compatible inward sweep.
The distinguished root \(r\) is the attachment point to the complement and may have several additional incident legs into that complement.
All physical legs on non-root vertices are inputs.
In the compatible inward sweep, every internal edge is oriented toward \(r\), so the non-root local unfoldings appear as coisometries.}
\label{fig:tree-root-sweep}
\end{figure}

\begin{definition}[One-sided rooted region]
Let \(\mathcal T\) be a connected TN.
Fix \(\sigma\in\{\mathrm{in},\mathrm{out}\}\).
A connected induced subnetwork \(R\subseteq \mathcal T\) with distinguished vertex \(r\in V(R)\) is called a \(\sigma\)-one-sided rooted region with root \(r\) if:
\begin{enumerate}
\item the internal bond graph induced by \(V(R)\) is a tree;
\item every internal bond edge of \(\mathcal T\) with exactly one endpoint in \(V(R)\) is incident to \(r\);
\item every physical leg attached to a vertex in \(V(R)\setminus\{r\}\) has type \(\sigma\).
\end{enumerate}
Thus non-root vertices of \(R\) have no bond connection to the complement, and all bond attachments of the region to the complement occur at the root.
The root may carry arbitrary physical legs.
The region is called \emph{maximal} if it is inclusion-maximal among rooted regions with the same type \(\sigma\).
\end{definition}

\begin{proposition}[Optimal preprocessing of one-sided rooted trees]\label{prop:io-tree-preprocess}
Let \(R\) be a one-sided rooted region with root \(r\).
After exact Schmidt-rank compression along the internal bonds of \(R\), there exists a compatible partial sweep through \(R\setminus\{r\}\), where parent and child are understood with respect to the tree rooted at \(r\).
For an input-one-sided region, the partial sweep is inward: every non-root vertex is processed before its parent, and the partial sweep is placed before the root step in any global sweep.
For an output-one-sided region, the partial sweep is outward: every parent is processed before its children, and the partial sweep is placed only after the root interface is already on the frontier.
With this compatibility convention:
\begin{enumerate}
\item every non-root local unfolding has local scale \(1\);
\item every output-one-sided non-root unfolding is an isometry and is unflagged under the default policy of Section~\ref{sec:local};
\item every input-one-sided non-root unfolding is a coisometry and therefore contributes no scale loss, although it is treated as a selected-block step under the default isometry-only shortcut; and
\item the partial contraction compiled by this sweep is scale-optimal.
\end{enumerate}

More explicitly:
\begin{itemize}
\item if \(R\) is input-one-sided, the contraction \(C_R\) of the non-root tensors of \(R\) is a coisometry onto the exact Schmidt support at the root interface and
\[
\Gamma_R=1=\|C_R\|_2;
\]
\item if \(R\) is output-one-sided, the contraction \(W_R\) of the non-root tensors of \(R\) is an isometry out of the exact Schmidt support at the root interface and
\[
\Gamma_R=1=\|W_R\|_2.
\]
\end{itemize}

Moreover, this preprocessing uses a number of arithmetic operations polynomial in the explicit input size of the rooted region.
\end{proposition}

\begin{proof}
Exact rooted canonicalization on a tree is obtained by successive exact Schmidt-rank compressions along the internal bonds together with the standard rooted gauge fixing that absorbs the nonisometric factors toward the root.
For each parent-facing bond, this replaces the bond space by the exact Schmidt support of the corresponding descendant subtree across that bond.

In the input-one-sided case, a sweep compatible with the rooted orientation processes every non-root vertex before its parent on the path to \(r\).
After exact Schmidt-rank compression, each parent-facing bond carries exactly the Schmidt support of the corresponding descendant subtree.
For a non-root vertex \(v\), the grouped local unfolding is therefore the map from the tensor product of its child-facing compressed bond spaces and local physical input legs to its compressed parent-facing bond.
By Lemma~\ref{lem:rooted-canonical}, this map is a coisometry.
Hence its local scale is \(1\).
Under the default isometry-only shortcut it is still realized as a selected-block step unless it is also an isometry, but this does not change the scale.

Composing these coisometries along the partial sweep yields a coisometry from the grouped descendant/input space onto the exact Schmidt support at the root interface.
This partial contraction is exactly \(C_R\).
Therefore
\[
\|C_R\|_2=1,
\qquad
\Gamma_R=1,
\]
so the partial sweep compiles \(C_R\) scale-optimally.

The output-one-sided case is dual, but its partial sweep is understood relative to a frontier that already contains the root interface.
Equivalently, in a global sweep the root-interface step has already occurred.
The compatible outward partial sweep then processes every parent before its children.
At each non-root vertex, the grouped local unfolding is the map from the compressed parent-facing bond to the tensor product of the child-facing compressed bond spaces and local physical output legs, and by Lemma~\ref{lem:rooted-canonical} this map is an isometry.
So every local scale is \(1\), and these steps are unflagged under the default policy.
Composing these maps yields an isometry \(W_R\) out of the exact Schmidt support at the root interface, and therefore
\[
\|W_R\|_2=1,
\qquad
\Gamma_R=1.
\]

The arithmetic-complexity statement is Lemma~\ref{lem:rooted-canonical}.
\end{proof}

\subsection{Bridge-hourglass forests}

We now specialize the preprocessing proposition to the hourglass geometry.

\begin{definition}[Bridge-hourglass network]
A connected TN is a \emph{bridge-hourglass network} if its internal bond graph can be written as
\[
T_{\mathrm{in}}\cup T_{\mathrm{out}},
\qquad
V(T_{\mathrm{in}})\cap V(T_{\mathrm{out}})=\{b\},
\]
where \(T_{\mathrm{in}}\) and \(T_{\mathrm{out}}\) are trees, all physical inputs are attached to vertices of \(T_{\mathrm{in}}\), and all physical outputs are attached to vertices of \(T_{\mathrm{out}}\).
\end{definition}

\begin{definition}[Bridge-hourglass forest]
A TN is a \emph{bridge-hourglass forest} if each connected component is a bridge-hourglass network.
\end{definition}

Figure~\ref{fig:hourglass} shows an example.
The input tree and output tree meet only at the bridge site \(b\), so after preprocessing the only remaining nontrivial scale contribution is at \(b\).

\begin{figure}[ht]
\centering
\begin{tikzpicture}[x=1mm,y=1mm,every node/.style={font=\small,inner sep=1pt}]


\node[dirtriangle left, minimum size=9mm] (l1) at (0,24) {};
\node[dirtriangle left, minimum size=9mm] (l2) at (0,0) {};
\node[dirtriangle left, minimum size=9mm] (l3) at (16,12) {};

\node[generaltensor, minimum width=8mm] (b) at (36,12) {\(b\)};

\node[dirtriangle right, minimum size=9mm] (r3) at (56,12) {};
\node[dirtriangle right, minimum size=9mm] (r1) at (72,24) {};
\node[dirtriangle right, minimum size=9mm] (r2) at (72,0) {};


\draw[sw_active] (l1) -- (l3);
\draw[sw_active] (l2) -- (l3);
\draw[sw_active] (l3) -- (b);

\draw[sw_active] (b) -- (r3);
\draw[sw_active] (r3) -- (r1);
\draw[sw_active] (r3) -- (r2);


\draw[dimleg] (-18,24) -- (l1);
\node[font=\scriptsize, anchor=east] at (-19,24) {in};

\draw[dimleg] (-18,0) -- (l2);
\node[font=\scriptsize, anchor=east] at (-19,0) {in};

\draw[dimleg] (-4,12) -- (l3);
\node[font=\scriptsize, anchor=east] at (-5,12) {in};

\draw[dimleg] (24,24) -- (b);
\node[font=\scriptsize, anchor=east] at (23,24) {in};


\draw[dimleg] (b) -- (48,2);
\node[font=\scriptsize, anchor=west] at (49,2) {out};

\draw[dimleg] (r3) -- (76,12);
\node[font=\scriptsize, anchor=west] at (77,12) {out};

\draw[dimleg] (r1) -- (90,24);
\node[font=\scriptsize, anchor=west] at (91,24) {out};

\draw[dimleg] (r2) -- (90,0);
\node[font=\scriptsize, anchor=west] at (91,0) {out};


\node[font=\small] at (10,36) {canonical input tree \(T_{\mathrm{in}}\)};
\node[font=\small] at (62,36) {canonical output tree \(T_{\mathrm{out}}\)};
\node[font=\small] at (36,30) {bridge site \(b\)};

\node[font=\scriptsize] at (36,-16)
{internal bond graph: \(T_{\mathrm{in}}\cup T_{\mathrm{out}}\), \(\;V(T_{\mathrm{in}})\cap V(T_{\mathrm{out}})=\{b\}\)};

\end{tikzpicture}
\caption{Bridge-hourglass network.
Arrows indicate the compatible sweep direction, while the triangle orientations indicate the logical coisometric/isometric orientation of the preprocessed one-sided trees.}
\label{fig:hourglass}
\end{figure}

\begin{proof}[Proof of Theorem~\ref{thm:main-hourglass}]
Consider one connected component of the forest.
Its input side is a maximal input-one-sided rooted region rooted at the bridge site \(b\), and its output side is a maximal output-one-sided rooted region rooted at the same site.

Apply Proposition~\ref{prop:io-tree-preprocess} to both rooted trees.
This exact preprocessing uses a number of arithmetic operations polynomial in the explicit input size and yields compatible partial sweeps on both sides that are scale-optimal.
Hence the represented operator factors as
\[
H(\mathcal T)=W_{\mathrm{out}}\,A_b\,C_{\mathrm{in}},
\]
where \(C_{\mathrm{in}}\) is a coisometry onto the exact input-side root Schmidt support, \(W_{\mathrm{out}}\) is an isometry out of the exact output-side root Schmidt support, and \(A_b\) is the unfolded operator at the bridge site on the compressed root support spaces.

Therefore
\[
\|H(\mathcal T)\|_2=\|A_b\|_2.
\]
If \(\|A_b\|_2=0\), then the bridge operator is zero, hence the global operator is zero, and we use Appendix~\ref{app:zero}.
Use the global sweep that processes the non-root input tree inward, then the bridge site \(b\), and then the non-root output tree outward.
On this sweep, every non-root site contributes local scale \(1\), by Proposition~\ref{prop:io-tree-preprocess}, and the bridge site contributes local scale \(\|A_b\|_2\).
Hence
\[
\Gamma(\pi)=\|A_b\|_2=\|H(\mathcal T)\|_2.
\]
So the sweep is scale-optimal.

For a bridge-hourglass forest, apply the same argument componentwise.
If any connected component preprocesses to the zero operator, then the full forest operator is zero, hence the result is scale-optimal by convention.
Otherwise all connected components are nonzero, the remaining operator is the tensor product of the single-site bridge operators of the connected components, and both operator norm and compiled scale multiply across tensor products.
Concatenating the componentwise scale-optimal sweeps again gives a scale-optimal sweep.
\end{proof}

\subsection{Complexity barrier for unrestricted exact preprocessing}

We now prove Proposition~\ref{prop:main-hardness} in the standard finite-encoding sense.
It is enough to consider integer-valued diagonal MPOs with binary-encoded entries.

\begin{proof}[Proof of Proposition~\ref{prop:main-hardness}]
Assume there exists a classical procedure running in time polynomial in the total bit length that, given a binary-encoded diagonal qubit MPO \(\mathcal M\) with local tensors in \(\mathrm{span}\{I,Z\}\), returns an equivalent TN representation, a sweep \(\pi\), and an exactly encoded value \(g\), comparable to integers in polynomial time, such that
\[
g=\Gamma(\pi)=\|H(\mathcal M)\|_2.
\]

Given an integer quadratic Ising instance
\[
E(z)=c+\sum_i l_i z_i+\sum_{i<j}\alpha_{ij}z_i z_j,
\qquad
c,l_i,\alpha_{ij}\in\Z,
\qquad
z\in\{\pm1\}^n,
\]
and an integer threshold \(T\), form the diagonal operator
\[
Q
=
cI
+
\sum_i l_i Z_i
+
\sum_{i<j}\alpha_{ij} Z_i Z_j
\]
and
\[
B:=|c|+\sum_i |l_i|+\sum_{i<j}|\alpha_{ij}|
\]
as in the standard integer Ising/QUBO correspondence \cite{Lucas2014}.
Then \(Q+BI\) is positive diagonal and
\[
\|Q+BI\|_2 = B+\max_z E(z).
\]

By the standard diagonal MPO construction, \(Q+BI\) has a polynomial-size qubit MPO representation with local tensors in \(\mathrm{span}\{I,Z\}\); see, e.g., \cite{Crosswhite2008,Pirvu2010}.
Applying the assumed procedure yields
\[
g=\|Q+BI\|_2.
\]
Hence
\[
\max_z E(z)\ge T
\iff
g-B\ge T.
\]
Since \(B\) and \(T\) are integers with polynomial bit length, the final comparison is polynomial time by assumption.
So threshold integer quadratic Ising optimization is solvable in polynomial time.

Therefore, unless \(\classP=\classNP\), no such unrestricted exact preprocessing theorem can exist even for diagonal MPOs on a path.
\end{proof}

\begin{remark}[Where the hardness lies]\label{rem:hardness-where}
For any fixed sweep \(\pi\), the compiled scale \(\Gamma(\pi)=\prod_{t}\beta_t\) is computable in polynomial time from the local unfoldings, so evaluating a given sweep is not the obstacle.
The barrier is recognizing optimality: a sweep is optimal exactly when \(\Gamma(\pi)=\|H(\mathcal T)\|_2\), and for a diagonal path MPO this norm is \(\|H(\mathcal T)\|_2=\max_z E(z)\), which is NP-hard to compute.
The difficulty therefore resides in the global operator norm, not in sweep search.
\end{remark}

\begin{remark}[No universal approximation ratio for the scale]\label{rem:no-approx}
The hardness barrier of Proposition~\ref{prop:main-hardness} concerns exact scale optimality.
A separate, stronger obstruction rules out any universal \emph{approximation ratio} for the scale achieved by a fixed sweep against the optimum: no such bound exists, already for two-site path TNs.

Consider a two-site path with input leg at the first site, output leg at the second, bond dimension \(2\), and local tensors \(\operatorname{diag}(1,a)\) and \(\operatorname{diag}(1,0)\) for \(a>1\).
The represented operator is
\[
\operatorname{diag}(1,a)\,\operatorname{diag}(1,0)=\operatorname{diag}(1,0),
\]
so \(\|H(\mathcal T)\|_2=1\) and the optimal scale is \(\Gamma_{\mathrm{opt}}=1\).
Under the left-to-right sweep, the local spectral scales are
\[
\beta_1=\|\operatorname{diag}(1,a)\|_2=a,
\qquad
\beta_2=\|\operatorname{diag}(1,0)\|_2=1,
\]
hence
\[
\Gamma(\pi)=a,
\qquad
\frac{\Gamma(\pi)}{\Gamma_{\mathrm{opt}}}=a.
\]
As \(a\to\infty\) the ratio is unbounded; substituting \(b=1/a>1\) into the same construction exhibits unboundedness as \(a\to 0\) as well.

Thus a fixed sweep can be arbitrarily far from scale-optimal, and no bound on \(\Gamma(\pi)/\Gamma_{\mathrm{opt}}\) in terms of the network graph, the local tensor norms, or any other sweep-independent quantity can exist.
This is the data-dependent counterpart to the combinatorial frontier-memory bound of Section~\ref{sec:complexity}: \(M(\pi)\) admits a graph-theoretic floor (pathwidth), but the scale cost does not, because it depends on cancellations in the tensor data that no graph parameter sees.
Exact preprocessing (Section~\ref{sec:structured}) restores optimality for the hourglass class; in general it cannot, by Proposition~\ref{prop:main-hardness}.
\end{remark}

\section{Discussion}\label{sec:discussion}

The practical output of the paper is not just a conversion from TNs to BEs, but an operator-level design interface for block-encoded quantum algorithms.
An explicit TN can be compiled directly to an explicit qubit BE without first reducing to MPO form, and the sweep records the exact scale, frontier memory, and dilation count paid by that choice of representation and layout.
The bounded-local correspondence and the selected-block round trip
\[
\mathrm{BE}\to\mathrm{TN}\to\mathrm{BE}
\]
show that TNs can also serve as a classical optimization layer for BE design.

\subsection{Selected-block optimization as a practical workflow}

A useful consequence of the round trip is that an explicit BE can be reduced to a TN for its selected block rather than for an arbitrary unitary extension.
If
\[
(\bra{0^a}\otimes I)\,U\,(\ket{0^a}\otimes I)=B,
\]
then fixing the ancilla preparation and post-selection boundary conditions turns the circuit into a TN for \(B\).

This makes the induced optimization problem operator-level.
One may therefore:
\begin{enumerate}
\item convert the explicit BE to a TN for its selected block,
\item optimize, compress, or approximate that TN classically, and
\item recompile the result to a new explicit BE.
\end{enumerate}
The guarantees depend sharply on which TN-side operation is performed (Remark~\ref{rem:round-trip-regimes}).
With no modification, recompilation is exact and scale-preserving.
Under faithful Schmidt-rank compression, Corollary~\ref{cor:round-trip-compress} gives a monotone guarantee: the scale cost and frontier memory cannot increase, and the operator error is bounded by the discarded Schmidt weight.
Arbitrary restructuring (gauge changes, reordering, or non-truncation edits) admits no general scale guarantee; this is unavoidable, since certifying exact scale optimality is already intractable for diagonal MPOs on a path (Proposition~\ref{prop:main-hardness}).
Thus the provable core of selected-block optimization is the identity and the faithful-compression regime; the heuristic regime is where practical TN manipulations live, without a general theorem.

This perspective is also useful in the forward direction.
One may choose a graph and a parameterized operator TN ansatz, optimize its local tensors as explicit classical data, and then compile the resulting operator to a BE.
The sweep quantities
\[
\Gamma(\pi),\quad M(\pi),\quad D(\pi)
\]
then provide explicit compiler-aware costs.
No trainability claim is made here, and poor scale control can still lead to poor post-selection behavior after compilation.

For structured families such as bridge-hourglass forests, the selected-block TN representation also exposes exact preprocessing opportunities before recompilation.
An explicit BE need not therefore be treated as a final form.

A third direction is the compression of composed BEs.
Given two BEs representing \(H_1=\alpha_1 B_1\) and \(H_2=\alpha_2 B_2\), their composition \(H_2H_1\) is itself a finite linear map whose circuit TN is obtained by concatenating the two circuit TNs, with each BE's preparation and post-selection ancillas imposed as fixed boundary values.
Every gate remains unitary, so the identity recompilation has \(\Gamma=1\) and scale \(\alpha_1\alpha_2\).
Applying faithful Schmidt-rank compression to the composition's TN and recompiling by Corollary~\ref{cor:round-trip-compress} yields a \emph{single} BE for \(H_2H_1\) with scale at most \(\alpha_1\alpha_2\).
When the composition has compressible structure (for instance, redundant ancilla bookkeeping or cancellable intermediate projections across the two stages), this collapses a sequence of separately post-selected BEs into one, potentially reducing the total post-selection cost.
No general reduction is guaranteed, since the bound is monotone rather than strict, but the capability is native to the interface: any operator built from BEs can be re-expressed as one compressed TN and recompiled as a single BE.

\subsection{Interface scope and remaining bottlenecks}

One need not first flatten the operator into a one-dimensional MPO representation, which may increase intermediate bond dimensions or obscure the graph structure that made the TN useful in the first place.
Thus states, effects, encoders, decoders, projections, and transfer operators can all be treated as native design objects.

The same native handling gives the interface a concrete role as a structured data-loading access model.
State preparation is the canonical instance: preparing \(\ket{\psi}\) is the linear map \(\C\to\mathcal H\), \(1\mapsto\ket{\psi}\), which is rectangular unless \(\dim\mathcal H=1\).
Native rectangular handling is therefore what lets state preparation be a BE of the state itself rather than an artificial square embedding with throwaway ancillas; the same applies to effects \(\mathcal H\to\C\) and to encoders and decoders between spaces of different dimension.
For operators and states admitting a compact TN description, the interface thus serves the data-loading function of a QRAM without a black-box oracle: the load is explicit, with frontier cost priced by pathwidth (Section~\ref{sec:complexity}) and post-selection cost by \(\Gamma(\pi)\).
The scope is strictly the structured subclass (data with compact TN descriptions), not arbitrary classical data, for which a black-box QRAM remains the only known access model and is outside the present scope.
Within the structured subclass, however, the interface replaces an unpriced black box with explicit, sweep-accounted circuits.

This does not remove graph-combinatorial difficulty.
Sweep memory is still controlled by
\[
M(\pi),
\]
and by Proposition~\ref{prop:cutwidth} this is exactly the weighted cutwidth of the augmented network graph under the chosen sweep.
So even before scale is considered, good compilation already requires a good layout.

Beyond the bounded-local regime, the main general statement is an interface statement:
every explicitly specified finite linear map can be represented as a TN and therefore compiled to an explicit BE with polynomial overhead in the explicit input size.
This is not a succinctness claim, and it is not an optimality claim.

In the bounded-local regime, the comparison becomes sharp.
Bounded-local TNs and bounded-local BEs correspond up to constant-factor overhead in the explicit model studied here.
Hence if an operator family does not admit polynomial-size bounded-local TNs, then it does not admit polynomial-size explicit bounded-local BEs in this model either.

From this viewpoint, classical TN choices such as graph structure, bond dimensions, gauges, symmetry sectors, and local compression schemes become part of BE architecture design.
The present results do not solve those design problems, but they make their effect on compilation explicit through the compiler scale, frontier memory, and dilation count.

\appendix

\section{Zero operators and degenerate scale gauge}\label{app:zero}

The only genuine degeneracy in the present model concerns the zero operator.
It appears if one allows BE with vanishing global scale,
\[
H=\alpha B,
\qquad
\alpha=0.
\]
In that case, when \(H=0\), the selected block \(B\) is no longer determined by the represented operator.
For example,
\[
0 = 1\cdot 0
\qquad\text{and}\qquad
0 = 0\cdot I
\]
are both valid formal factorizations, but they induce different selected-branch semantics.
In the first case the selected block is zero, whereas in the second it is the identity.
Thus the degeneracy is not that the represented operator vanishes, but that allowing \(\alpha=0\) destroys uniqueness of the selected block.

For this reason, the main text adopts the nondegenerate convention
\[
\alpha>0.
\]
This does \emph{not} exclude zero operators from the theory.
It only excludes the additional zero-gauge convention with vanishing global scale.

A separate issue is local normalization.
If an unfolded site operator satisfies
\[
\beta_v=\|A^{(v)}\|_2=0,
\]
then \(A^{(v)}=0\), so every full contraction of the TN vanishes and therefore
\[
H(\mathcal T)=0.
\]
Thus \(\beta_v=0\) is an immediate local certificate for the zero operator.
The compiler may then terminate early and return the canonical zero-operator output
\[
\alpha=1,
\qquad
\widehat H=0.
\]
For example, add one ancilla initialized in \(\ket{0}\), apply \(X\) to it, and post-select it in \(\ket{0}\).

The converse need not hold.
A TN may satisfy
\[
H(\mathcal T)=0
\]
even though every local unfolding has positive norm.
For instance,
\[
\operatorname{diag}(1,0)\,\operatorname{diag}(0,1)=0.
\]
So local zero unfoldings detect only a special subclass of zero operators.
They are a shortcut, not a characterization.

Accordingly, the compiler distinguishes two cases:

\begin{enumerate}
\item If some local unfolding has \(\beta_v=0\), then zero is certified locally and the compiler may immediately output the canonical zero BE with \(\alpha=1\) and selected block \(0\).
\item If all local unfoldings satisfy \(\beta_v>0\), then the main compilation flow is well defined.
This still allows the final operator to vanish globally.
In that case one simply obtains
\[
H(\mathcal T)=0
\qquad\text{with}\qquad
\Gamma(\pi)>0
\]
and the compiled selected block is
\[
\widehat H_\pi=0.
\]
\end{enumerate}

No efficient general global zero-test is claimed here.
The only special handling used by the compiler is the local certificate \(\beta_v=0\).

\section{General-dimension padding}\label{app:padding}

In the main text, every local leg dimension is first padded to a power of two.
Thus the local Hilbert spaces already satisfy the atomic qubit-register convention used by the compiler, the memory accounting, and certified-zero reuse.

There is also a purely abstract finite-dimensional variant which avoids padding every leg separately.
It is only a tensor-factor bookkeeping observation and is not used elsewhere in the paper.

Let
\[
A\in\C^{m\times n},
\qquad
\beta=\|A\|_2>0,
\qquad
C:=\beta^{-1}A .
\]
At the matrix level, any square padding dimension
\[
k\ge \max(m,n)
\]
is enough to form a square contraction
\[
\widetilde C=J_{\mathrm{out}}CJ_{\mathrm{in}}^\dagger
\]
and then apply the same one-flag dilation construction.

If one wants the same abstract ambient space to be compatible with both the input-side and output-side tensor factorizations, a canonical choice is
\[
k:=\lcm(m,n).
\]
Then
\[
\C^k
\cong
\C^n\otimes \C^{k/n}
\cong
\C^m\otimes \C^{k/m}.
\]
Thus the rectangular map \(C:\C^n\to\C^m\) may be viewed as the selected block of a square contraction on a common abstract \(k\)-dimensional register, with different input-side and output-side factorizations.
Since
\[
k\le mn,
\]
this changes local dimensions by at most a quadratic factor in the explicit local dimensions.

If a qubit implementation is ultimately required, the abstract \(k\)-dimensional register must still be embedded into a power-of-two Hilbert space.
This observation therefore does not replace the atomic qubit-register convention of the main text, and it does not by itself give the certified-zero qubit slots or the qubit memory accounting used there.

\section{Proof of Proposition~\ref{prop:lipschitz}}\label{app:lipschitz-proof}

\begin{proof}
Introduce the hybrid sequence
\[
C^{(t)}
:=
\Phi(A_1,\dots,A_t,\widetilde A_{t+1},\dots,\widetilde A_L),
\qquad
t=0,\dots,L,
\]
so that
\[
C^{(L)}=H(\mathcal T),
\qquad
C^{(0)}=H(\widetilde{\mathcal T}).
\]
Then
\[
H(\mathcal T)-H(\widetilde{\mathcal T})
=
\sum_{t=1}^{L}\left(C^{(t)}-C^{(t-1)}\right).
\]

Fix \(t\).
With the sweep order held fixed, contract the first \(t-1\) sites into the left frontier map \(L_t\)
and the sites \(t+1,\dots,L\) of the hybrid network into the right frontier map \(R_t\).
Then the \(t\)-th hybrid difference factors as
\[
C^{(t)}-C^{(t-1)}
=
R_t\,(A_t-\widetilde A_t)\,L_t.
\]
Therefore, by submultiplicativity of the operator norm,
\[
\|C^{(t)}-C^{(t-1)}\|_2
\le
\|R_t\|_2\,\|A_t-\widetilde A_t\|_2\,\|L_t\|_2.
\]

It remains to bound \(\|L_t\|_2\) and \(\|R_t\|_2\).
Each of these maps is itself obtained by composing the corresponding local unfolded site operators along the chosen sweep through the intermediate frontier spaces.
Hence repeated submultiplicativity gives
\[
\|L_t\|_2
\le
\prod_{j<t}\|A_j\|_2
=
\prod_{j<t}\beta_j,
\]
and
\[
\|R_t\|_2
\le
\prod_{j>t}\|\widetilde A_j\|_2
=
\prod_{j>t}\widetilde\beta_j.
\]
So
\[
\|C^{(t)}-C^{(t-1)}\|_2
\le
\left(
\prod_{j<t}\beta_j
\right)
\|A_t-\widetilde A_t\|_2
\left(
\prod_{j>t}\widetilde\beta_j
\right).
\]
Summing over \(t\) proves the first claim.

If \(\beta_t,\widetilde\beta_t\le M\) for all \(t\), then every product above is bounded by \(M^{L-1}\), and therefore
\[
\bigl\|H(\mathcal T)-H(\widetilde{\mathcal T})\bigr\|_2
\le
M^{L-1}\sum_{t=1}^{L}\|A_t-\widetilde A_t\|_2.
\]
\end{proof}

\section{A worked example: sweep-dependent scale and frontier memory}\label{app:toy-example}

This appendix records a small instance, deferred from Section~\ref{sec:complexity}, illustrating how the sweep changes the compiled scale \(\Gamma(\pi)\) and frontier memory \(M(\pi)\) for a fixed operator, and how exact TN-side contraction restores a scale loss that a bad sweep had introduced.

Consider a two-site path \(v_1-v_2\) with one global input leg at \(v_1\), one global output leg at \(v_2\), bond dimension \(2\), and local tensors
\[
T^{(v_1)}=\operatorname{diag}(1,0),
\qquad
T^{(v_2)}=I_2.
\]
Then
\[
H(\mathcal T)=|0\rangle\!\langle 0|.
\]

For the sweep \((v_1,v_2)\), the induced local maps are the obvious input-to-bond projector at \(v_1\) and bond-to-output identity at \(v_2\).
Hence the local scales are both \(1\), so
\[
\Gamma(v_1,v_2)=1.
\]
Under the default local realization policy, the first step is genuinely dilated while the second is unflagged, and under one-qubit leg padding the frontier width is always \(1\).
Thus
\[
D(v_1,v_2)=1,
\qquad
M(v_1,v_2)=1,
\qquad
\rho(v_1,v_2)=1.
\]

For the reverse sweep \((v_2,v_1)\), the same tensors are unfolded instead as the unnormalized map
\[
1\mapsto |00\rangle_{b,o}+|11\rangle_{b,o}
\]
at \(v_2\) and an effect
\[
\langle 00|_{i,b}
\]
at \(v_1\), whose composition is again \(|0\rangle\!\langle 0|\).
The first step therefore has local scale \(\sqrt2\) and, after normalization, is an isometry; the second has local scale \(1\) and is a coisometry.
Under the default isometry-only shortcut, the coisometric effect is still realized by the one-flag selected-block primitive.
Hence
\[
\Gamma(v_2,v_1)=\sqrt2,
\qquad
D(v_2,v_1)=1.
\]
Under one-qubit leg padding, after the first step the frontier simultaneously carries the global input leg, the internal bond, and the global output leg, so
\[
M(v_2,v_1)=3,
\qquad
\rho(v_2,v_1)=\frac{1}{\sqrt2}.
\]

Thus the represented operator is unchanged, but the sweep changes the scale and frontier memory:
the forward sweep is scale-optimal and uses smaller frontier memory, while the reverse sweep incurs worse scale.

The reverse sweep is not optimal, but it becomes so under a trivial classical preprocessing step.
Contracting the two sites of the same network into a single site gives a one-site TN representing the same operator \(|0\rangle\!\langle 0|\).
Compiling that one-site TN gives a single local step with local scale \(1\), hence
\[
\Gamma=1,
\qquad M=1,
\qquad D\in\{0,1\},
\qquad \rho=1,
\]
recovering scale-optimality with no post-selection loss.
In the round-trip language of Section~\ref{sec:discussion}, this is an exact TN-side contraction of the selected-block TN of an existing BE, followed by recompilation; the transfer results of Section~\ref{sec:complexity} make the improvement carry over to the recompiled BE.
This is the smallest instance of the general phenomenon that exact TN preprocessing acts at the operator level before compilation, and it already shows the round trip restoring a scale loss that a bad sweep had introduced.

\printbibliography

\end{document}